\documentclass[journal,final]{ieeetran}
\usepackage{graphicx}
\ifCLASSOPTIONcompsoc
\usepackage[caption=false,font=normalsize,labelfont=sf,textfont=sf]{subfig}
\else
\usepackage[caption=false,font=footnotesize]{subfig}
\fi
\usepackage{multirow}
\usepackage{array}
\usepackage[noadjust]{cite}
\usepackage{url}
\usepackage{diagbox}
\usepackage{mathtools}
\usepackage{amsmath}
\usepackage{amssymb}
\usepackage{empheq}
\usepackage{bm}
\usepackage{color}
\usepackage{gensymb}
\usepackage{algorithm}
\usepackage{algpseudocode}
\usepackage{amsthm}
\usepackage[many]{tcolorbox}
\usepackage[shortlabels]{enumitem}
\usepackage{booktabs}

\usepackage[makeroom]{cancel}

\newcounter{mytempeqncnt}

\makeatletter
\def\cantox@vector#1#2#3#4#5#6#7#8{%
  \dimen@.5\p@
  \setbox\z@\vbox{\boxmaxdepth.5\p@
   \hbox{\kern-1.2\p@\kern#1\dimen@$#7{#8}\m@th$}}%
  \ifx\canto@fil\hidewidth  \wd\z@\z@ \else \kern-#6\unitlength \fi
  \ooalign{%
    \canto@fil$\m@th \CancelColor
    \vcenter{\hbox{\dimen@#6\unitlength \kern\dimen@
      \multiply\dimen@#4\divide\dimen@#3 \vrule\@depth\dimen@\@width\z@
      \vector(#3,-#4){#5}%
    }}_{\raise-#2\dimen@\copy\z@\kern-\scriptspace}$%
    \canto@fil \cr
    \hfil \box\@tempboxa \kern\wd\z@ \hfil \cr}}
\def\bcancelto#1#2{\let\canto@vector\cantox@vector\cancelto{#1}{#2}}
\makeatother

\makeatletter
\newcommand*\rel@kern[1]{\kern#1\dimexpr\macc@kerna}
\newcommand*\widebar[1]{%
  \begingroup
  \def\mathaccent##1##2{%
    \rel@kern{0.8}%
    \overline{\rel@kern{-0.8}\macc@nucleus\rel@kern{0.2}}%
    \rel@kern{-0.2}%
  }%
  \macc@depth\@ne
  \let\math@bgroup\@empty \let\math@egroup\macc@set@skewchar
  \mathsurround\z@ \frozen@everymath{\mathgroup\macc@group\relax}%
  \macc@set@skewchar\relax
  \let\mathaccentV\macc@nested@a
  \macc@nested@a\relax111{#1}%
  \endgroup
}
\makeatother

\theoremstyle{definition}

\newtcolorbox{framefloat}[1][]{fonttitle=\bfseries, titlerule=0pt, boxrule=0.5pt,
  colframe=black,colback=white,float=!t,#1}
  
\newcommand\blfootnote[1]{%
  \begingroup
  \renewcommand\thefootnote{}\footnote{#1}%
  \addtocounter{footnote}{-1}%
  \endgroup
}

\makeatletter
\newcommand\fs@spaceruled{\def\@fs@cfont{\bfseries}\let\@fs@capt\floatc@ruled
  \def\@fs@pre{\vspace{1\baselineskip}\hrule height.8pt depth0pt \kern2pt}%
  \def\@fs@post{\kern2pt\hrule\relax}%
  \def\@fs@mid{\kern2pt\hrule\kern2pt}%
  \let\@fs@iftopcapt\iftrue}
\makeatother

\begin{document}
\title{Hybrid RIS With Sub-Connected Active Partitions: Performance Analysis and Transmission Design}
\author{Konstantinos Ntougias,~\textit{Member, IEEE}, Symeon Chatzinotas,~\textit{Fellow, IEEE}, and Ioannis Krikidis,~\textit{Fellow, IEEE}}

\maketitle

\begin{abstract}
The emerging reconfigurable intelligent surface (RIS) technology promises to enhance network capacity via passive reflect beamforming. However, product path loss limits its performance gains. Fully-connected (FC) active RIS integrates a reflect-type power amplifier into each element to address this issue, while sub-connected (SC) active RIS, which employs partitioning and power amplifier sharing per partition, and hybrid FC-active/passive RIS utilize fewer power amplifiers to tackle the resulting cost and energy consumption challenges. This study introduces novel hybrid RIS structures with at least one SC-active reflecting sub-surface (RS) to achieve a better capacity/energy efficiency (EE) balance. A system model considering power amplifier sharing and resulting in a proper diagonal SC-active RIS/RS beamforming matrix is derived for the first time. The asymptotic signal-to-noise-ratio of active/passive and active/active RIS with a single shared power amplifier per active RS in a single-user single-input single-output setup is analyzed. The transmit and RIS beamforming weights are jointly optimized to maximize the EE of a hybrid RIS-aided multi-user multiple-input single-output downlink system under the power consumption constraints of the base station and the RIS. Numerical results highlight the substantial performance gains of the proposed designs over benchmarks and provide insights.
\end{abstract}

\begin{IEEEkeywords}
Reconfigurable intelligent surface (RIS), sub-connected RIS, hybrid RIS, performance analysis, optimization.
\end{IEEEkeywords}

\section{Introduction}\label{sec:1}
\blfootnote{K. Ntougias and I. Krikidis are with the Department of Electrical and Computer Engineering, University of Cyprus, 1678 Nicosia, Cyprus. S. Chatzinotas is with SnT, University of Luxembourg, 1855 Luxembourg City, Luxembourg and with College of Electronics \& Information, Kyung Hee University, Yongin-si, 17104, Korea. (E-mail: ntougias.konstantinos@ucy.ac.cy, symeon.chatzinotas@uni.lu, krikidis@ucy.ac.cy).}
The disruptive reconfigurable intelligent surface (RIS) technology has been recently proposed as a means to increase the capacity of wireless communication systems in a cost-effective and energy-efficient manner. Specifically, by jointly adjusting the phase shifts that its passive reflecting elements independently induce to the incident radio frequency (RF) signal, this planar surface can reflect that signal as desired, e.g., constructively to boost the signal power at intended users or/and destructively to suppress the interference in other directions~\cite{IRS2,IRSSurvey}. However, the large, due to its multiplicative nature, path loss over the RIS-cascaded channels between the base station (BS) and the users renders these channels weak and, therefore, limits the relative capacity gain. This performance deterioration is particularly pronounced when the respective direct BS--user channels are strong~\cite{IRSEE}.

A fully-connected (FC) active RIS alternative has been recently introduced to alleviate this performance bottleneck~\cite{IRSAct1,IRSAct2}. In this RIS variant, a low-cost reflection-type power amplifier, which is realized with the aid of an active load (negative resistance) such as a tunnel diode, is integrated into each RIS element. By amplifying the incident RF signal prior to reflection, FC-active RIS effectively addresses the so-called product path loss issue mentioned above.

Nevertheless, since each reflect-type power amplifier requires power supply, the large-scale deployment of FC-active RIS nodes raises cost and power consumption considerations. Hence, it is important to reduce the total power consumption (TPC) by using fewer reflect-type power amplifiers than RIS elements. At the same time, it is desirable to provide flexible control of and strike a beneficial capacity/energy efficiency (EE) balance. A sub-connected (SC) active RIS structure, wherein each reflect-type power amplifier feeds its own RIS partition~\cite{Subconnected}, and a hybrid RIS architecture that combines a FC-active and a passive reflecting sub-surface (RS)~\cite{HybridActPasRIS}, constitute recent advancements under this spirit.

Note that while passive RIS provides only phase shift control per element, active RIS variants offer also amplification control, either per element (FC) or per partition (SC). Thus, by properly adjusting the number of partitions or active elements in SC-active or FC-active/passive RIS, we can correspondingly reduce the TPC and achieve a respective capacity/EE trade-off.

\subsection{Related Works}\label{subsec:1.1}
Numerous studies have explored joint transmit precoding/RIS beamforming optimization in a variety of scenarios. 

\textbf{Passive RIS:}~\cite{Provable} addresses transmit sum-power (TSP) minimization in a multi-user multiple-input single-output (MISO) downlink setting, while~\cite{IRS3,IRSImpCSI1} extend this work considering discrete reflection coefficients (RC) or imperfect knowledge of the RIS—user channels, respectively. Sum-rate (SR) and EE maximization are explored in~\cite{IRSPwr, IRSEE}, respectively, under the application of zero-forcing precoding. Weighted sum-rate (WSR) maximization is investigated in~\cite{IRSWSRmax}. Secrecy rate maximization in single-user MISO or multiple-input multiple-output (MIMO) systems with single- or multi-antenna eavesdroppers is tackled in~\cite{IRSPhySec1,IRSPhySec3}, respectively.~\cite{WCL2} considers the multi-objective optimization problem of maximizing the SR and the sum-harvested-energy in multi-user MISO downlink systems with discrete information and energy receivers.~\cite{IRSSWIPT0,IRSSWIPT1} deal with the maximization of the minimum power or the weighted sum-power at the energy receivers, and~\cite{IRSSWIPT2} handles TSP minimization.~\cite{IRSSWIPTImpCSI, NtougiasProb, NtougiasHPIRS} revisit the latter problem for an equivalent setting with power-splitting receivers under imperfect channel state information (CSI), a spectrum underlay regime, or the application of hybrid precoding, respectively.~\cite{CB1,CB2} consider joint channel estimation and codebook-based RIS beamforming optimization to reduce optimization complexity and signaling overhead. 

These works corroborate the gains of RIS-aided communication and optimal phase shifts over benchmarks, such as full-duplex active relays and random phase shifts, respectively.

\textbf{FC-active RIS:} In~\cite{IRSAct4}, the authors jointly optimize the transmit precoding and RIS beamforming schemes in a multi-user MISO downlink system to minimize the TSP. The work~\cite{IRSAct8} focuses instead on SR maximization, whereas the study~\cite{CFRISEEF} presents an iterative algorithm that maximizes EE fairness in a cell-free network. In~\cite{IRSAct11}, the authors minimize the TSP in a single-user MISO system collocated with a single-antenna eavesdropper, while~\cite{IRSAct12} considers secrecy-rate maximization. The work~\cite{IRSAct13} maximizes the worst-case secrecy rate or the weighted sum-secrecy-rate in a MISO broadcast channel with a single-antenna eavesdropper. In~\cite{IRSAct10}, the authors maximize the WSR in a wireless powered communication network, whereas in~\cite{IRSAct5} they minimize the TSP in a MISO broadcast channel with power-splitting receivers. Other works are concerned with the maximization of the weighted sum-power at the energy receivers or the WSR of the information receivers~\cite{IRSAct6,IRSAct7}. 

According to the studies mentioned above, active RIS outperforms its passive counterpart under the same TPC budget, provided that the number of RIS elements (and, therefore, reflect-type power amplifiers) is relatively small, due to TPC and amplification noise considerations.

\textbf{SC-active and FC-active/passive RIS:}~\cite{Subconnected} focuses on EE maximization in a SC-active RIS-aided multi-user MISO downlink system. Numerical simulations unveil that this active RIS variant provides significant TPC savings with minimal SR deterioration, compared to FC-active RIS. However, this study ignored the inevitable power combination and re-distribution when multiple phase shift control circuits are connected to a single (shared) amplification control circuit. This issue is addressed in~\cite{SC3}. Assuming equal power allocation among the elements in each RIS partition, the authors tackle the SR maximization and TSP minimization problems, confirming the TPC savings and performance improvement under limited TPC budget offered by SC-active RIS compared to FC-active RIS. Nevertheless, their model doesn't result in a diagonal RIS beamforming matrix, thus failing to highlight the connection with other RIS structures, and complicates optimization. The authors in~\cite{HybridActPasRIS} also optimize RIS elements' allocation in a FC-active/passive RIS-assisted MISO broadcast channel to maximize the ergodic capacity of the worst-case user under the availability of statistical CSI. Numerical evaluations indicate that smaller TPC budgets and stronger direct links benefit active over passive RIS elements. The works~\cite{SC2, SC1} consider a corresponding unmanned aerial vehicle-aided communication setup. The authors tackle max-min fairness and optimize the location or trajectory of the hovering or moving drone, respectively.~\cite{CFHybRISEE,FCP2} demonstrate the EE gains of this hybrid RIS design when used as reflecting relay or transmitting reflectarray over passive and active RIS in a cell-free network or MISO broadcast channel, respectively. 

\subsection{Motivation, Goal, and Contributions}\label{subsec:1.2}
Wireless networks should reduce their TPC and carbon footprint and enhance their EE, in view of the energy crisis, the climate change, and their densification, which is driven by their ambitious capacity targets~\cite{GreenDeal}. Motivated by the SC-active and FC-active/passive RIS variants, which represent steps in this direction, this work advances the state-of-the-art by introducing novel RIS designs combining the hybrid and SC-active RIS principles, to further improve the SR/EE balance as well as to enhance our ability to control it under various objectives, constraints, and operating conditions.

In particular, we introduce a new active/passive and two active/active RIS structures, namely, SC-active/passive, FC-active/SC-active, and SC-active/SC-active (also called hybrid SC-active) RIS, respectively. Compared to the aforementioned RIS designs, the proposed hybrid RIS architectures potentially present the following advantages:
\begin{enumerate}
\item \textbf{Cost-effectiveness:} SC-active/passive and SC-active/SC-active RIS designs are more cost-effective than FC-active and FC-active/passive structures, due to the smaller number of reflect-type power amplifiers, while offering a good balance between performance and hardware costs, as indicated by the pure SC-active RIS structure. Thus, they represent a better fit for large-scale deployments.
\item \textbf{Energy efficiency:} The proposed architectures allow for fine-grained control over the performance--power consumption trade-off, by adjusting the size (and, therefore, number) of SC-active RIS partitions as in SC-active/passive and SC-active/SC-active RIS, possibly along with the number of active elements that are fed each by their own reflect-type power amplifier, as in FC-active/SC-active RIS. This is crucial for meeting the increasing demand for green communications.
\item \textbf{Scalability:} The employment of SC-active reflecting subsurfaces (RS) facilitates easier scaling of the RIS size, as adding more elements does not necessarily require a proportional increase in the number of power amplifiers.
\item \textbf{Flexibility:} The proposed hybrid RIS architectures provide additional design degrees-of-freedom and enable adaptation to various deployment scenarios and performance requirements, thus offering more options in the middle ground between the two extreme cases of purely passive and fully-active RIS implementations. For instance, SC-active/SC-active RIS essentially corresponds to an SC-active RIS with \textit{non-uniform} partitioning, whereas FC-active/SC-active RIS combines per element and per group amplification control to support a desirable performance/energy consumption trade-off.
\end{enumerate}

After performing asymptotic SNR analysis to better understand the behavior of the proposed hybrid RIS designs compared to the state-of-the-art, we focus on optimal system design under an EE maximization objective, motivated by the increasing importance of sustainable and green communications in contemporary and next-generation wireless networks, as mentioned earlier. EE is a particularly important performance metric for hybrid RIS architectures, where the interplay between passive and active elements (with individual or shared reflect-type power amplifiers) introduces new dimensions to the power-performance trade-off. Although EE maximization has been studied for active and hybrid FC-active/passive RIS systems, our work breaks new ground by addressing the unique challenges posed by the introduced SC-active/passive and hybrid active/active architectures. Specifically, we tackle the complex interplay between FC-active, SC-active, and passive structures within a single RIS, considering their distinct power consumption models and performance characteristics, thereby providing insights into the optimal balance between different types of RIS elements and structures in hybrid designs.

Furthermore, system modeling efficiently incorporates power amplifier sharing in SC-active RSs, an aspect not considered in the majority of earlier studies such as~\cite{Subconnected} and~\cite{IRSAct8}, while maintaining tractability, contrary to similar efforts in the literature such as~\cite{SC3}. The novel mathematical model introduced in this work streamlines optimization, enabling the reuse of standard, simple to implement and well-studied optimization techniques, and facilitates direct comparisons with other RIS architectures and works.

The unique contributions of this study are listed below: 
\begin{itemize} 
\item We introduce new active/passive and active/active hybrid RIS structures employing the SC-active design. 
\item We propose, for the first time to the best of our knowledge, a system model for SC-active RIS/RS accounting for power amplifier sharing and, in contrast to~\cite{SC3}, resulting in a diagonal RIS beamforming matrix, thus highlighting the connection with FC-active RIS/RS while also substantially simplifying optimization compared to~\cite{SC3}. 
\item We analyze the asymptotic SNR of hybrid RIS designs with a single shared reflect-type amplifier per active RS in a single-input single-output (SISO) system as the number of elements grows to infinity, considering the resulting power combination and re-distribution, as opposed to the asymptotic SNR derivations in~\cite{IRSAct8}. 
\item Considering these hybrid RIS types with fixed elements' allocation, we jointly optimize the transmit precoding and RIS beamforming schemes to maximize the EE in a RIS-aided multi-user MISO downlink system, subject to the TPC constraints of the BS and the RIS. We apply fractional programming (FP) and develop efficient block coordinate ascent (BCA) algorithms using the Lagrange multipliers and majorization-minimization (MM) methods to tackle these challenging non-convex optimization tasks and obtain high-quality solutions. 
\end{itemize} 

Numerical simulation results in various test scenarios show the significant performance gains of the proposed designs against benchmarks and provide valuable insights.

\subsection{Structure and Mathematical Notation}
The rest of the paper is organized as follows: The considered hybrid RIS structures and the system model are introduced in Section~\ref{sec:2}. Section~\ref{sec:3} is devoted to performance analysis. In Section~\ref{sec:4}, the joint transmit precoding/RIS beamforming optimization problems for each scenario are formulated, and respective energy-efficient designs are developed. In Section~\ref{sec:5}, the performance of the proposed schemes is comparatively evaluated against benchmarks via numerical simulations. Section~\ref{sec:6} summarizes this work and discusses future extensions.

\textbf{Notation:} $x$: scalar; $\mathbf{x}$: column vector; $\mathbf{X}$: matrix; $\left[\mathbf{x}\right]_{n}\triangleq x_{n}$: $n$-th element of $\mathbf{x}$; $\mathbb{R}_{+}$: set of non-negative real numbers; $\mathbb{C}^{N}$: set of complex $N$-dimensional vectors; $\mathbb{B}^{N\times M}$: set of binary $N\times M$ matrices; $\left|x\right|$, $\operatorname{arg}(x)$, and $\operatorname{Re}\{x\}$: magnitude, argument, and real part of $x\in\mathbb{C}$; $\left\|\mathbf{x}\right\|$: Euclidean norm of $\mathbf{x}$; $\mathbf{X}^{*}$, $\mathbf{X}^{T}$, $\mathbf{X}^{\dagger}$, $\mathbf{X}^{-1}$, $\mathbf{X}^{+}$, $\left\|\mathbf{X}\right\|_{F}$, and $\operatorname{Tr}\left(\mathbf{X}\right)$: complex conjugate, transpose, complex conjugate transpose, inverse, Moore-Penrose pseudo-inverse, Frobenius norm, and trace of $\mathbf{X}$; $\mathbf{X}=\operatorname{diag}\left(\mathbf{x}\right)$: diagonal matrix with main diagonal $\mathbf{x}=\operatorname{Diag}\left(\mathbf{X}\right)$; $\mathbf{X}=\operatorname{blkdiag}\left(\mathbf{X}_{1},\dots,\mathbf{X}_{L}\right)$: block diagonal matrix with blocks $\mathbf{X}_l$, $l=1,\dots,L$; $\mathbf{X}\succeq \mathbf{0}$: positive semi-definite (PSD) matrix; $\lambda_{\max}\left(\mathbf{X}\right)$: maximum eigenvalue of $\mathbf{X}$; $\mathbf{0}_{N}$ and $\mathbf{1}_{N}$: $N$-dimensional null and unity vectors; $\mathbf{I}_{N}$: $N\times N$ identity matrix; $\otimes$ and $\odot$: Kronecker and Handamard matrix product; $\mathbb{E}\left\{\cdot\right\}$: expectation operator; $\mathcal{CN}\left(\cdot,\cdot\right)$: circularly symmetric complex Gaussian distribution.
\begin{figure}[!t]
    \centering
    \includegraphics[scale = 0.6]{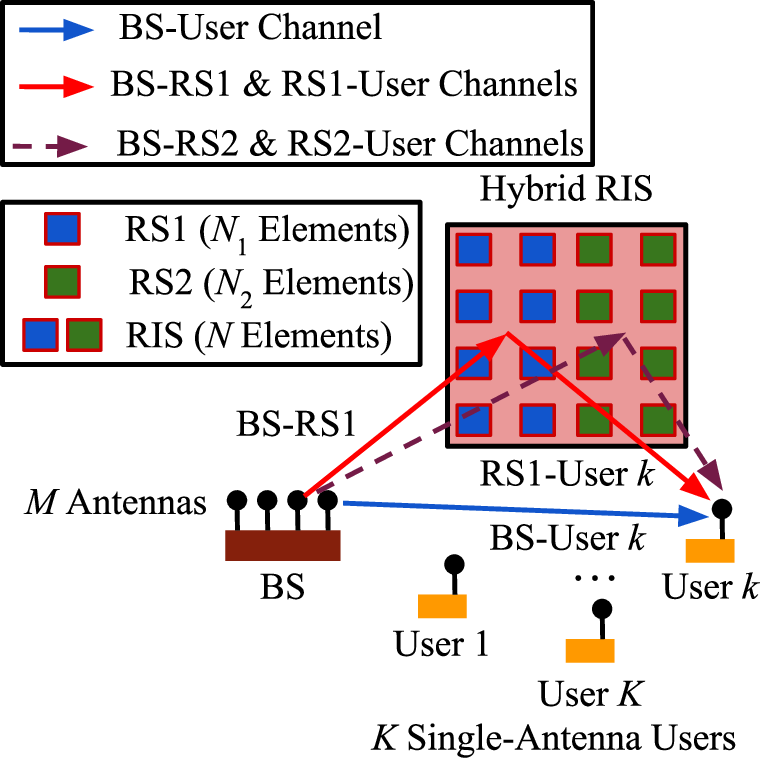}
    \caption{System setup: A BS equipped with $M$ antennas serves $K$ single-antenna users on a single time-frequency resource with the aid of a hybrid RIS having $N$ elements. The RIS is partitioned into two RSs, namely, RS1 and RS2, with $N_1$ and $N_2$ elements, respectively.}
    \label{fig:1}
\end{figure}

\section{Hybrid RIS Structures and System Model}\label{sec:2}

\subsection{System Setup}\label{subsec:2.1}
The system consists of a BS and a hybrid RIS that jointly serve a set $\mathcal{K}\triangleq\left\{1,\dots,K\right\}$ of single-antenna terminals on a single time-frequency resource, as shown in Fig.~\ref{fig:1}. The BS is equipped with $M$ antennas. The hybrid RIS has $N$ elements split among two RSs, i.e., RS $s$ is equipped with a set $\mathcal{N}_s$ of $N_s$ elements, $s\in\mathcal{S}\triangleq\{1,2\}$, with $N_1=aN$ and $N_{2}=(1-a)N$, $a\in(0,1)$, such that $N_1+N_2=N$. RS $s$ employs the passive, FC-active, or SC-active design. In the latter case, it is divided into a set $\mathcal{L}_{s}\triangleq\left\{1,\dots,L_{s}\right\}$ of partitions with $T_{s}=N_{s}/L_{s}$ elements each. We address the members of these sets via $n\in\mathcal{N}_{s}$, $l\in\mathcal{L}_{s}$, and $k,i\in\mathcal{K}$, respectively.

\subsection{Hybrid RIS Architectures and RIS Beamforming Matrix}\label{subsec:2.2}
\begin{figure}[!t]
    \centering
    \includegraphics[width=\columnwidth]{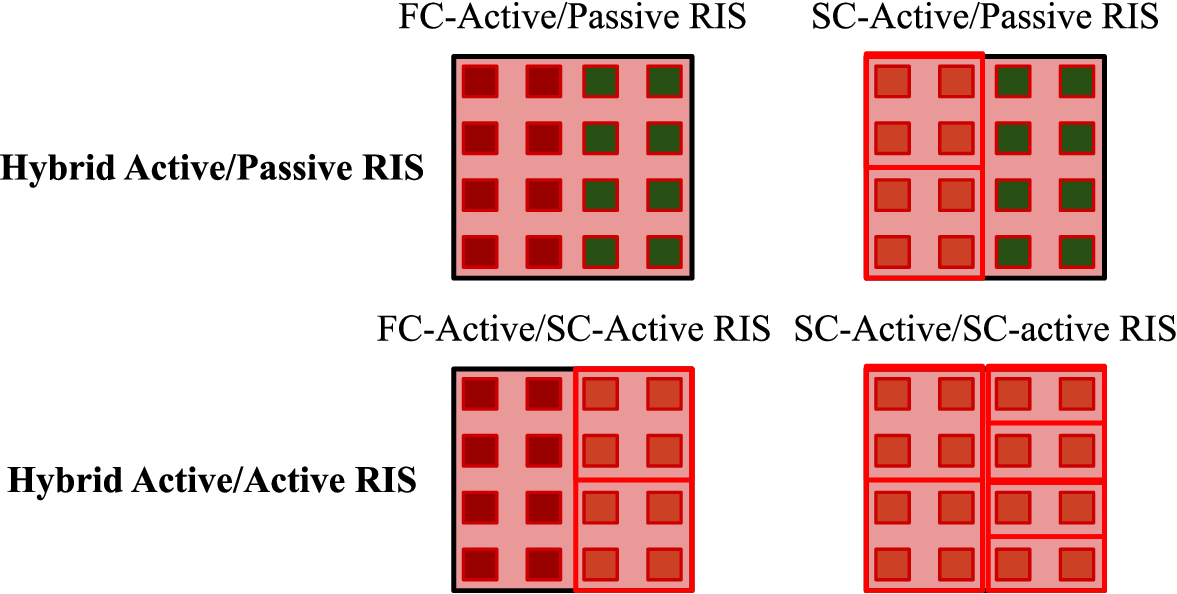}
    \caption{Hybrid RIS architectures. In this example, each RIS has $N = 16$ elements divided into two RSs with $N_1=N_2=8$ elements each. Thus, in FC-active/passive RIS, there are $8$ power amplifiers, each one feeding its own element. In SC-active/passive RIS, in turn, we note that the SC-active RS is partitioned into 2 groups with $4$ elements each. Hence, this RIS structure utilizes $2$ power amplifiers, with each one being shared among $4$ elements. Likewise, FC-active/SC-active RIS makes use of $8+2 = 10$ power amplifiers. Finally, in SC-active/SC-active RIS, RS2 employs $4$ groups of $2$ elements each, thereby resulting in $2+4 = 6$ power amplifiers. This example shows that the various hybrid RIS architectures differ in the number of power amplifiers and the level where amplification control is applied.}
    \label{fig:1c}
\end{figure}

We consider active/passive and active/active RIS architectures, i.e., the FC-active/passive RIS design and the proposed SC-active/passive, FC-active/SC-active, and hybrid SC-active RIS structures, as illustrated in Fig.~\ref{fig:1c}. The latter involves non-uniform partitioning, i.e., $L_{1}\neq L_{2}$ or/and $N_{1}\neq N_{2}$. The hybrid RIS beamforming matrix, $\bm{\Phi}\in\mathbb{C}^{N\times N}$, is $\bm{\Phi}=\operatorname{blkdiag}\left(\bm{\Phi}_{1},\bm{\Phi}_{2}\right)$, where $\bm{\Phi}_{s}\in\mathbb{C}^{N_{s}\times N_{s}}$ is the beamforming matrix of RS $s$. Generally, $\bm{\Phi}_{s}=\operatorname{diag}\left(\bm{\phi}_s\right)=\operatorname{diag}\left(\phi_{s,1},\dots,\phi_{s,N_s}\right)=\mathbf{B}_s\bm{\Theta}_s$, where $\phi_{s,n}=\beta_{s,n}e^{j\theta_{s,n}}$, $\beta_{s,n} = \left|\phi_{s,n}\right| \geq 0$, $\theta_{s,n}=\operatorname{arg}\left(\phi_{s,n}\right)\in[0,2\pi)$, and $j\triangleq\sqrt{-1}$ denote the RC of element $n$ in RS $s$, its amplitude and phase shift, and the imaginary unit, respectively, while $\mathbf{B}_s=\operatorname{diag}\left(\bm{\beta}_s\right)=\operatorname{diag}\left(\beta_{s,1},\dots,\beta_{s,N_s}\right)\in\mathbb{R}_{+}^{N_s\times N_s}$ and $\bm{\Theta}_s=\operatorname{diag}\left(e^{j\theta_{s,1}},\dots,e^{j\theta_{s,N_s}}\right)\in\mathbb{C}^{N_s\times N_s}$ respectively represent the amplitudes and phase shifts matrix. In the FC-active structure, $\beta_{s,n}\leq \beta_{\max}$, $\beta_{\max}>1$, whereas in the passive one, $\beta_{s,n}=1$, $\forall n\in\mathcal{N}_s$, i.e., $\bm{\beta}_s=\mathbf{1}_{N_s}\Leftrightarrow\mathbf{B}_s=\mathbf{I}_{N_s}$. In the SC-active design, assuming equal power allocation among the elements in each partition, $\mathbf{B}_s=\frac{1}{\sqrt{T_s}}\operatorname{diag}\left(\bm{\Gamma}_s\widetilde{\bm{\beta}}_s\right)$ and $\bm{\Theta}_s = \widetilde{\bm{\Theta}}_s\widetilde{\bm{\Theta}}_s^T$, where the coupling matrix $\bm{\Gamma}_{s}\in\mathbb{B}^{N_{s}\times L_{s}}$ is defined as $\bm{\Gamma}_{s}\triangleq\mathbf{I}_{L_{s}}\otimes\mathbf{1}_{T_{s}}$, $\widetilde{\bm{\beta}}_s\triangleq\left[\widetilde{\beta}_{s,1},\dots,\widetilde{\beta}_{s,L_s}\right]^T\in\mathbb{R}_{+}^{L_s}$, $\widetilde{\beta}_{s,l}\leq\beta_{\max}$ denotes the common amplitude of the elements in partition $l$ of RS $s$, and $\widetilde{\bm{\Theta}}_s\triangleq\operatorname{diag}\left(e^{j\widetilde{\theta}_{s,1}},\dots,e^{j\widetilde{\theta}_{s,N_s}}\right)\in\mathbb{C}^{N_s\times N_s}$, with $\widetilde{\theta}_{s,n}\in[0,2\pi)$. That is, $\bm{\Phi}_s = \operatorname{blkdiag}\left(\bm{\Phi}_{s,1},\dots,\bm{\Phi}_{s,L_s}\right)$, where $\bm{\Phi}_{s,l}\in\mathbb{C}^{T_s\times T_s}$ is the RCs matrix of partition $l$, $\bm{\Phi}_{s,l}\triangleq\frac{1}{\sqrt{T_s}}\widetilde{\beta}_{s,l}\bm{\Theta}_{s,l}$, and $\bm{\Theta}_{s,l}\in\mathbb{C}^{T_s\times T_s}$ denotes the phase shifts matrix of this partition, $\bm{\Theta}_{s,l}\triangleq\operatorname{diag}\left(e^{j\bm{\theta}_{s,l}}\right)$, with $\bm{\theta}_{s,l}\triangleq\left[\theta_{s,(l-1)T_s+1},\dots,\theta_{s,lT_s}\right]^T\in\mathbb{R}^{T_s}$. If $T_s=1$, such that $L_s=N_s$, $\bm{\Gamma}_s=\mathbf{I}_{N_s}$, $\widetilde{\bm{\beta}}_s=\bm{\beta}_s$, and $\bm{\Phi}_{s,l}\rightarrow\phi_{s,n}$, then $\bm{\Phi}_{\text{sc}} = \bm{\Phi}_{fc}\widetilde{\bm{\Theta}}^T$. Hence, in general, $\bm{\Theta}_s = \operatorname{diag}\left(e^{j\operatorname{arg}\left(\bm{\phi}_{s}\right)}\right)$ and $\bm{\beta}_s=\bm{\Gamma}_s^{+}\operatorname{diag}\left(e^{-j\operatorname{arg}\left(\bm{\phi}_{s}\right)}\right)\bm{\phi}_{s}$, while $\widetilde{\theta}_{s,n}=\theta_{s,n}/2$.

\begin{figure}[!t]
    \centering
    \includegraphics[width=\columnwidth]{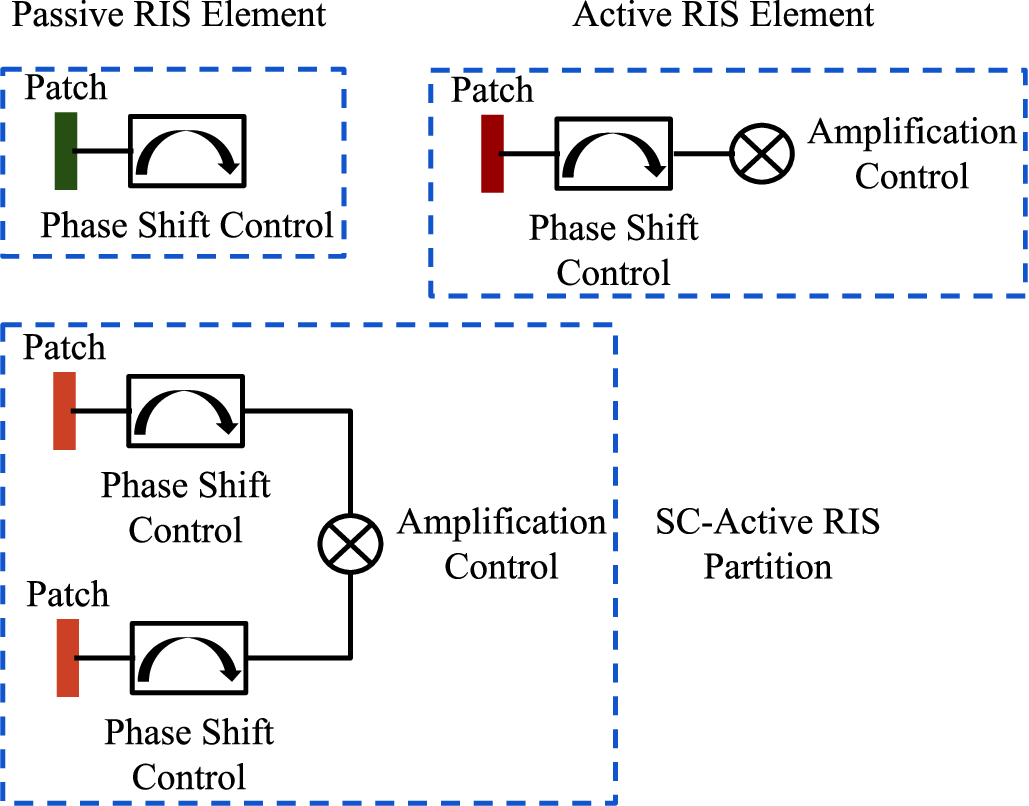}
    \caption{Each RS independently adopts the passive, FC-active, or SC-active RIS architecture. All these RIS structures provide phase shift control per element. The active RIS variants provide also amplification control. Specifically, FC-active RIS offers amplification control per element, since each element is connected to its own reflect-type power amplifier. SC-active RIS, on the other hand, enables amplification control per partition, since each group of elements shares a common power amplifier. This implies that each element in FC-active RIS imposes its own phase shift and amplitude to the incident RF signal, while the elements in an SC-active RIS partition induce different phase shifts but the same amplitude to it. Furthermore, from a circuit implementation perspective, we note that in an SC-active RIS partition, each element (patch) individually phase shifts the incident RF signal. Then, the outputs of the phase shift control circuits are added together and this sum is amplified. In the reverse signal path towards the patch, where the signal will be eventually reflected to the direction dictated by the phase shifts, the combined signal power is inevitably re-distributed to the RIS elements in the partition.}    
    \label{fig:1b}
\end{figure}

Note that the necessity of incorporating power amplifier sharing in our model is dictated by the inner mechanics of the SC-active implementation itself, as shown in Fig.~\ref{fig:1b}. Our formulation of the SC-active RS beamforming matrix explicitly accounts for power amplifier sharing, contrary to previous works focusing on performance analysis~\cite{IRSAct8} or optimization~\cite{Subconnected}, by: 1) using a common amplitude $\tilde{\beta}_{s,l}$ for all elements in partition $l$, reflecting the shared amplification, and utilizing the coupling matrix $\Gamma_s$ to represent the connectivity between shared amplifiers and RIS elements; 2) incorporating the $\frac{1}{\sqrt{T_s}}$ factor to model uniform power allocation among the $T_s$ elements in each partition; and 3) incorporating signal re-distribution via the term $\tilde{\Theta}_s \tilde{\Theta}_s^T$. The only other work, to the best of our knowledge, that captures in system modeling the inevitable power combination (forward path in the patch/RIS element) and re-distribution (reverse path) when multiple phase shift control circuits are connected to a single (shared) amplification control circuit is~\cite{SC3}. However, the model proposed in~\cite{SC3} is more complicated, since it directly maps the hardware structure and signal path, thus resulting in a non-diagonal RIS beamforming matrix and non-trivial mathematical expressions. This complexity is reflected into the optimization process as well. In this work, we propose a simpler model that accurately captures the practical constraints of power amplifier sharing while maintaining a tractable optimization framework. Specifically, this model ``hides'' the inner-complexity of the RIS structure, thus allowing us to apply standard optimization methods as in simpler RIS architectures. Finally, our model provides greater insights on the operation of SC-active and relevant hybrid RIS structures and highlights the connection between the mathematical representations of different RIS variants in a natural manner, as another advantage. In summary, we provide a common mathematical framework for modeling and optimizing any type of RIS structure, unveiling the similarities and differences between RIS variants and simplifying optimization.

When a fixed hybrid RIS structure is considered, the proposed designs do not bring along any additional complexity. Note that, in terms of implementation, an SC-active RIS partition consists of $L-1$ passive elements and one active element. The phase shifted signals from the passive elements are added together and the sum is added with the output of the phase shifter of the active element. This final sum then passes through the power amplifier of the active element. When RIS elements scheduling is considered, the use of switches (or diodes that act as switches) is required to dynamically connect or disconnect the power amplifiers in each element, as in FC-active/passive RIS, or connect adders for the summation of phase-shifted signals, depending on RIS elements allocation.

\subsection{Channel Model}\label{subsec:2.3}
We consider quasi-static, flat-fading channels. The baseband equivalent direct channel from the BS to user $k$, incident channel from the BS to RS $s$, reflected channel from RS $s$ to user $k$, and effective channel from the BS to user $k$ is denoted by $\mathbf{g}_{k}^{\dagger}\in\mathbb{C}^{M}$, $\mathbf{G}_{s}\in\mathbb{C}^{N_{s}\times M}$, $\mathbf{f}_{k,s}^{\dagger}\in\mathbb{C}^{N_{s}}$, and $\mathbf{h}_{k}^{\dagger}\in\mathbb{C}^{M}$, respectively, where $\mathbf{h}_{k}^{\dagger}\triangleq\mathbf{g}_{k}^{\dagger}+\sum_{s\in\mathcal{S}}\mathbf{f}_{k,s}^{\dagger}\bm{\Phi}_{s}\mathbf{G}_{s}$.

\subsection{Transmitted and Received Signals, TPC, and SINR}\label{subsec:2.4}
The transmitted signal, $\mathbf{x}\in\mathbb{C}^{M}$, is $\mathbf{x} = \sum_{k\in\mathcal{K}}\mathbf{w}_{k}s_{k}$, where $\mathbf{w}_{k}\in\mathbb{C}^{M}$ and $s_{k}\sim\mathcal{CN}(0,1)$ are the precoding vector and data symbol, respectively, for user $k$. The TSP is
\begin{equation}\label{eq:1}
    P_{t} = \mathbb{E}\left\{\left\|\mathbf{x}\right\|^{2}\right\} = \sum_{k\in\mathcal{K}}\left\|\mathbf{w}_{k}\right\|^{2}.
\end{equation}

Depending on whether RS $s$ is active or passive, the amplified and reflected or simply reflected signal, $\mathbf{t}_{s}\in\mathbb{C}^{N_{s}}$, is written as $\mathbf{t}_{s}=\bm{\Phi}_{s}\left(\mathbf{r}_{s}+\mathbf{z}_{s}\right)$ or $\mathbf{t}_{s}=\bm{\Phi}_{s}\mathbf{r}_{s}$, respectively, where $\mathbf{r}_{s}\in\mathbb{C}^{N_{s}}$ denotes the incident RF signal given by $\mathbf{r}_{s}=\mathbf{G}_{s}\mathbf{x}$ and $\mathbf{z}_{s}\sim\mathcal{CN}\left(\mathbf{0}_{N_{s}},\delta_{s}^{2}\mathbf{I}_{N_{s}}\right)$ represents active RS's amplification noise. The reflect power of active RS $s$ is
\begin{equation}\label{eq:2}
    P_{r,s} = \mathbb{E}\left\{\left\|\mathbf{t}_{s}\right\|^{2}\right\} = \sum_{k\in\mathcal{K}}\left\|\bm{\Phi}_{s}\mathbf{G}_{s}\mathbf{w}_{k}\right\|^{2} + \delta_{s}^{2}\left\|\bm{\Phi}_{s}\right\|_{F}^{2}.
\end{equation}

The TPC of the BS is written as $P_{\text{BS}} = \xi^{-1}P_{t} + W_{\text{BS}}$ and that of RS $s$ is given by $P_{s}=N_{s}P_{\text{PS}}$ if it is passive or $P_{s} = \zeta_{s}^{-1}P_{r,s}+W_{r,s}$ if it is active, where $W_{\text{BS}}$ and $W_{r,s}$ respectively refer to the static power consumption of the BS and active RS $s$, $\xi\in(0,1)$ and $\zeta_{s}\in(0,1)$ is the EE of their amplifiers, and $P_{\text{PS}}$ denotes the power consumption of each phase shift control circuit. $W_{r,s}=N_{s}\left(P_{\text{PS}}+P_{\text{DC}}\right)$ if active RS $s$ adopts the FC architecture or $ W_{r,s}=N_{s}P_{\text{PS}}+\left(N_{s}/T_{s}\right)P_{\text{DC}}$ if it employs the SC one, where $P_{\text{DC}}$ denotes the direct current (DC) bias of each power amplifier and $N_{s}/T_{s}=L_{s}$. The TPC constraint of the BS and active RS $s$ is given by $P_{\text{BS}} \leq P_{\text{BS}}^{\max}$ and $P_{s} \leq P_{s}^{\max}$, respectively, where $P_{\text{BS}}^{\max}>0$ and $P_{s}^{\max}>0$ denote the corresponding TPC budgets. By denoting $P_{\text{RIS}}\triangleq\sum_{s\in\mathcal{S}}P_{s}$, we can write the TPC of the system as $P=P_{\text{BS}}+P_{\text{RIS}}$.

The received signal at user $k$ is given by $y_k = \mathbf{g}_{k}^{\dagger}\mathbf{x}+\sum_{s\in\mathcal{S}}\mathbf{f}_{k,s}^{\dagger}\mathbf{t}_{s}+n_{k}$, where $n_{k}\sim\mathcal{CN}\left(0,\sigma_{k}^{2}\right)$ denotes the additive white Gaussian noise (AWGN). When an active/passive RIS is employed, the signal-to-interference-plus-noise ratio (SINR) of user $k$ is written as
\begin{equation}\label{eq:3}
    \gamma_{k} = \frac{\left\|\mathbf{h}_{k}^{\dagger}\mathbf{w}_{k}\right\|^{2}}{\sum\limits_{i\in\mathcal{K}\setminus\{k\}}\left\|\mathbf{h}_{k}^{\dagger}\mathbf{w}_{i}\right\|^{2}+\delta_{1}^{2}\left\|\mathbf{f}_{k,1}^{\dagger}\bm{\Phi}_{1}\right\|^{2}+\sigma_{k}^{2}}.
\end{equation}
When an active/active RIS is utilized, the SINR of user $k$ is expressed as
\begin{equation}\label{eq:4}
    \gamma_{k} = \frac{\left\|\mathbf{h}_{k}^{\dagger}\mathbf{w}_{k}\right\|^{2}}{\sum\limits_{i\in\mathcal{K}\setminus\{k\}}\left\|\mathbf{h}_{k}^{\dagger}\mathbf{w}_{i}\right\|^{2}+\sum\limits_{s\in\mathcal{S}}\delta_{s}^{2}\left\|\mathbf{f}_{k,s}^{\dagger}\bm{\Phi}_{s}\right\|^{2}+\sigma_{k}^{2}}.
\end{equation}
\begin{figure*}[!t]
\normalsize
\setcounter{mytempeqncnt}{\value{equation}}
\setcounter{equation}{4}
\begin{align}\label{eq:5}
\gamma_{a/p}\left(N,P_{t,a/p}^{\max},P_{r,1}^{\max}\right) &\rightarrow \frac{P_{t,a/p}^{\max}\pi^{2}\left[P_{r,1}^{\max}\varrho_{f_{1}}^{2}\varrho_{g_{1}}^{2}N_{1}+4\varrho_{f_{2}}^{2}\varrho_{g_{2}}^{2}\left(P_{t,a/p}^{\max}\varrho_{g_{1}}^{2}+\delta_{1}^{2}\right)N_{2}^{2}\right]}{16\left(P_{r,1}^{\max}\delta_{1}^{2}\varrho_{f_{1}}^{2}+4P_{t,a/p}^{\max}\sigma^{2}\varrho_{g_{1}}^{2}+4\sigma^{2}\delta_{1}^{2}\right)} \nonumber \\
&=\gamma_{a}\left(N_{1},P_{t,a/p}^{\max},P_{r,1}^{\max}\right) +\gamma_{p}\left(N_{2},P_{t,a/p}^{\max}\right)\frac{4\left(P_{t, a / p}^{\max }\varrho_{g_1}^2+\delta_1^2\right) \sigma^2}{P_{r, 1}^{\max } \delta_1^2 \varrho_{f_1}^2+4P_{t, a / p}^{\max } \sigma^2 \varrho_{g_1}^2+4\sigma^2 \delta_1^2}.
\end{align}
\setcounter{equation}{\value{mytempeqncnt}}
\hrulefill
\vspace*{4pt}
\end{figure*}

\subsection{Sum-Rate and Energy Efficiency}\label{subsec:2.7}
The data rate of user $k$ and the SR (in bps/Hz) are respectively given by $R_{k}=\log_{2}\left(1+\gamma_{k}\right)$ and $R=\sum_{k\in\mathcal{K}}R_{k}$, while the EE of the system is written as $\eta=R/P$.
\begin{figure*}[!t]
\centering
\subfloat[]{
	\label{subfig:2a}
	\includegraphics[scale = 0.3]{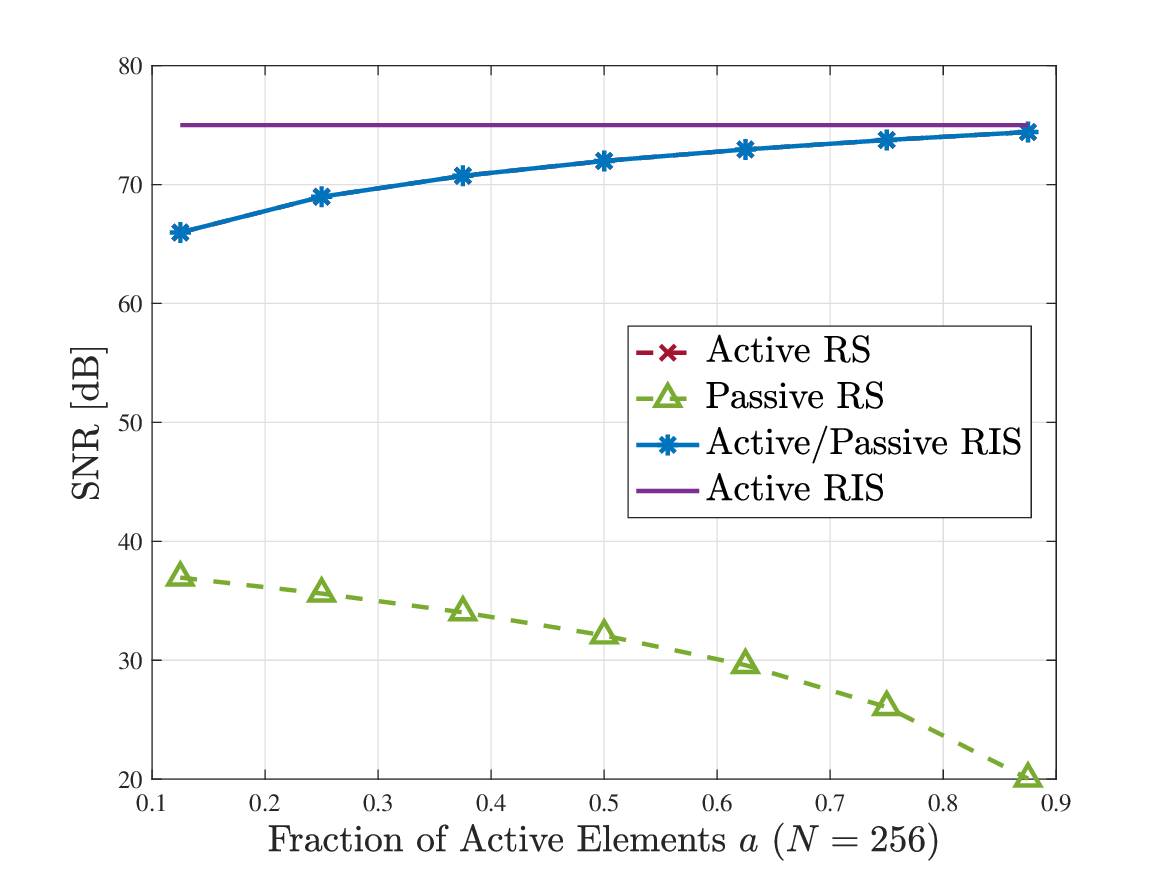} } 
\subfloat[]{
	\label{subfig:2b}
	\includegraphics[scale = 0.3]{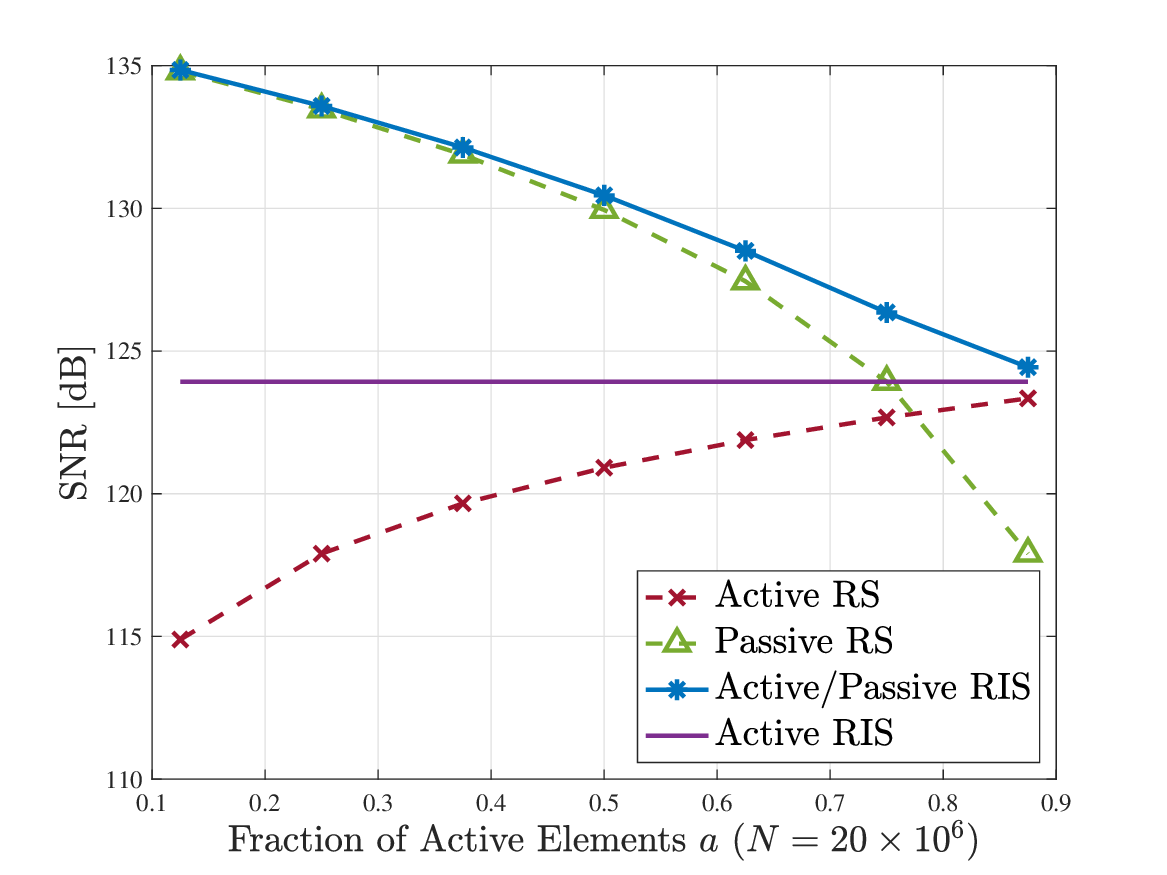} }
\subfloat[]{
	\label{subfig:2c}
	\includegraphics[scale = 0.3]{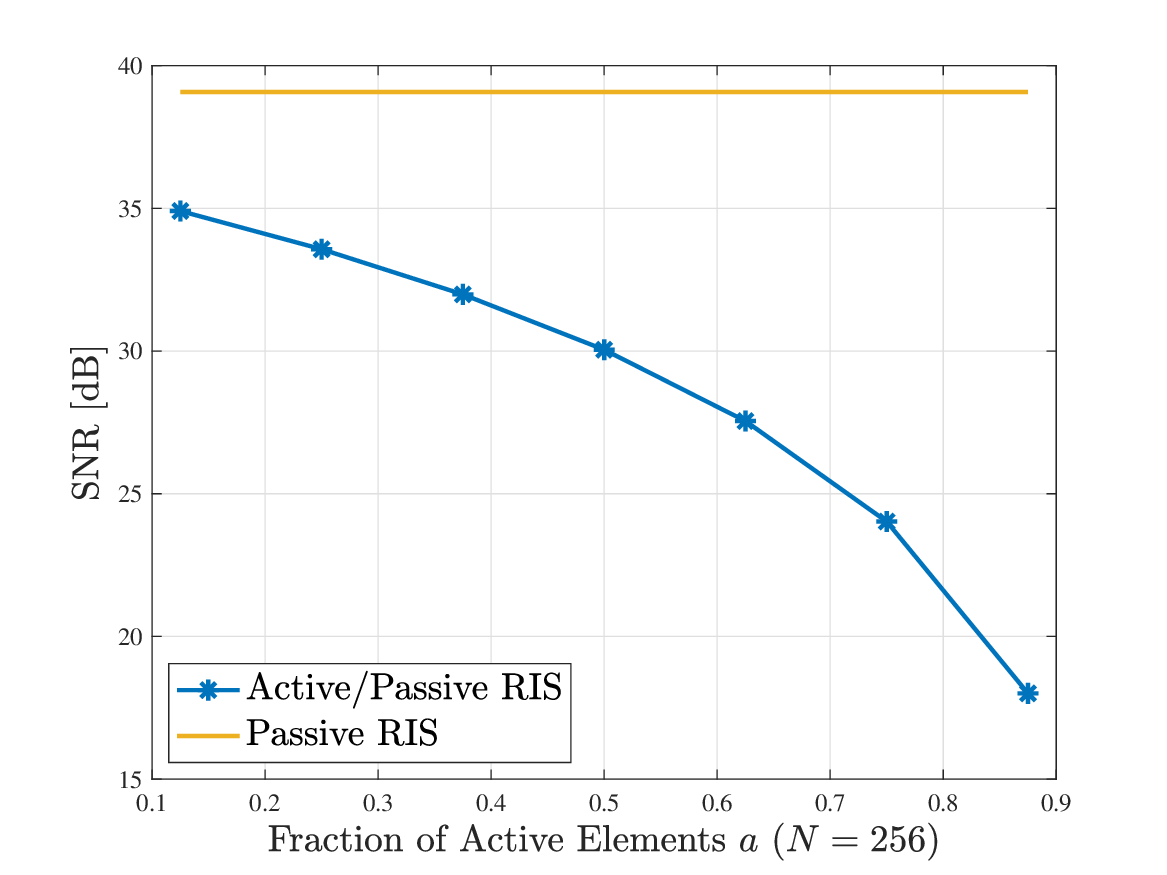} }  
\caption{SNR of active/passive vs. active or passive RIS as we vary the fraction of active elements in the former, for different operation regimes: (a) standard regime, (b) large number of RIS elements, and (c) large transmit power budget.}
\label{fig:2}
\end{figure*}

\section{Performance Analysis}\label{sec:3}
In this section, we analyze the asymptotic SNR of the considered hybrid RIS architectures, as $N_s\rightarrow\infty$. We adopt the common practice of considering a SISO setup ($M=K=1$) with Rayleigh-fading BS--RS $s$ and RS $s$--user channels, $\mathbf{g}_{s}\sim\mathcal{CN}\left(\mathbf{0}_{N_{s}},\varrho_{g_{s}}^{2}\mathbf{I}_{N_{s}}\right)$ and $\mathbf{f}_{s}\sim\mathcal{CN}\left(\mathbf{0}_{N_{s}},\varrho_{f_{s}}^{2}\mathbf{I}_{N_{s}}\right)$, respectively, and blocked direct BS--user link, to foster mathematical tractability and obtain insights about the respective power scaling laws~\cite{IRS1,IRSAct8,Subconnected}. For the same purpose, we also assume that if RS $s$ is active, then the amplitudes of its elements are equal. This corresponds to an SC-active RS with a single power amplifier (i.e., $L_s=1$), such that $\widetilde{\beta}_{l,s}=\beta_{s}$, $\forall l\in\mathcal{L}_{s}$. This configuration has been proven to be optimal for single-user SISO setups~\cite{Subconnected}. Since $L_s=1$, increasing the number of RSs in active/active RIS, $S$, is equivalent to increasing the number of partitions in SC-active RIS, $L$. The received signal is given by $y=h^{*}x+n$, where $h^{*}\triangleq\sum_{s\in\mathcal{S}}\mathbf{f}_{s}^{\dagger}\bm{\Phi}_{s}\mathbf{g}_{s}$ is the effective BS--user channel, $n\sim\mathcal{CN}\left(0,\sigma^2\right)$ represents the AWGN, and $x=ws$ denotes the transmitted signal, with $w$ and $s\sim\mathcal{CN}(0,1)$ standing for the transmit power and the transmitted symbol, respectively. Optimal transmit power and RCs are assumed. For comparison purposes, we also consider the asymptotic SNR of passive and active RIS as $N\rightarrow\infty$. 

We assume that the total radiated power budget, $P_{\max}$, is the same in all setups, to ensure fair comparison. Hence, we have for the transmit and reflect power budgets: $P_{t,p}^{\max}=P_{\max}$ (passive RIS), $P_{t,a}^{\max}+P_{r}^{\max}=P_{\max}$ (active RIS), $P_{t,a/p}^{\max}+P_{r,1}^{\max}=P_{\max}$ (active/passive RIS), and $P_{t,a/a}^{\max}+\sum_{s\in\mathcal{S}}P_{r,s}^{\max}=P_{\max}$ (active/active RIS). To enhance our insights, we also consider the special case of equal total radiated power budget splitting: $P_{t,p}^{\max} = P_{\max}$, $P_{t,a}^{\max}= P_{t,a/p}^{\max} = P_{t,a/a}^{\max} =P_{\max}/2$, $P_{r}^{\max}=P_{r,1}^{\max}=P_{\max}/2$, $P_{r,s}^{\max}=P_{\max}/2S=P_r^{\max}/S$. 

\subsection{Asymptotic SNR}\label{subsec:3.1}
\textbf{Lemma 1 (Asymptotic SNR for active/passive RIS):} The SNR for arbitrary elements' allocation is given by Eq.~(\ref{eq:5}) at the top of this page, where
\begin{equation*}
\gamma_{a}\left(N_1,P_{t,a/p}^{\max},P_{r,1}^{\max}\right)\rightarrow
\end{equation*}
\addtocounter{equation}{1}
\begin{align}\label{eq:7}
& N_1\frac{P_{t,a/p}^{\max}P_{r,1}^{\max}\pi^{2}\varrho_{f_1}^{2}\varrho_{g_1}^{2}}{16\left(P_{r,1}^{\max}\delta_1^{2}\varrho_{f_1}^{2}+4P_{t,a/p}^{\max}\sigma^{2}\varrho_{g_1}^{2}+4\sigma^{2}\delta_1^{2}\right)} \nonumber \\
&=aN\frac{P_{\max}^2 \pi^2 \varrho_{f_1}^2 \varrho_{g_1}^2}{32\left(P_{\max}\delta_1^2\varrho_{f_1}^2 + 4P_{\max}\sigma^2\varrho_{g_1}^2+8\sigma^2\delta_1^2\right)},
\end{align}
denotes the asymptotic SNR of the active RS (RS1), and
\begin{align}\label{eq:9}
    \gamma_{p}\left(N_2,P_{t,a/p}^{\max}\right) &\rightarrow  N_2^{2}\frac{P_{t,a/p}^{\max}\pi^{2}\varrho_{f_2}^{2}\varrho_{g_2}^{2}}{16\sigma^{2}} \nonumber \\
    &= (1-a)^2N^2\frac{P_{\max} \pi^2 \varrho_{f_2}^2 \varrho_{g_2}^2}{32\sigma^2},
\end{align}
represents the asymptotic SNR of the passive RS (RS2).

\textit{Proof:} Please see Appendix~\ref{App:A}. \qed

\textbf{Remark 1:} We note in Eq.~(\ref{eq:5}) that, under equal split, $\gamma_{a/p}\left(N,P_{\max}/2,P_{\max}/2\right)=\gamma_{a}\left(aN,P_{\max}/2,P_{\max}/2\right) +c\gamma_{p}\left((1-a)N,P_{\max}\right)$, i.e., the asymptotic SNR of an active/passive RIS with $N$ elements equals the sum of the asymptotic SNR of an active RIS having $aN < N$ elements with that of a passive RIS having $(1-a)N < N$ elements under $P_{\max}/2 < P_{\max}$ transmit power budget, where the latter SNR is multiplied by a constant $c < 1$. 

We consider a running numeric example throughout this section with the following parameters~\cite{IRSAct8}: $\varrho_{g_s}^2=\varrho_g^2=-70$ dB, $\varrho_{f_s}^2=\varrho_f^2 = -70$ dB, and $\delta_s^2=\delta^2=-100$ dBm, $\forall s\in\mathcal{S}$; $\sigma^2=-100$ dBm, and $P_{\max}=2$ W. Notice that for these values, $c=0.8$. In Fig.~\ref{subfig:2a}, we plot the SNR of active/passive RIS, $\gamma_{a/p}$, as we vary the fraction of active elements, $a=0.125:0.125:0.875$, versus that of active RIS, assuming in both cases $N=256$ elements. We also plot the SNR contribution of the active RS, $\gamma_{a}$, and the passive RS, $c\gamma_p$. We note that the SNR of the latter is from about 29 dB when $a=0.125$  up to 57.37 dB when $a=0.875$ lower than the SNR of the former. Consequently, the SNR of the active RS essentially coincides with that of the active/passive RIS, i.e., the SNR impact of the passive RS is negligible. Therefore,
\begin{align}\label{eq:11}
\gamma_{a/p}\left(N,\frac{P_{\max}}{2},\frac{P_{\max}}{2}\right) &\approx \gamma_{a}\left(aN,\frac{P_{\max}}{2},\frac{P_{\max}}{2}\right) \nonumber \\
&= a\gamma_{a}\left(N,\frac{P_{\max}}{2},\frac{P_{\max}}{2}\right).
\end{align}  
Hence, since $a<1$, the active/passive RIS with $N$ elements presents an SNR loss of $a$, or $10\log_{10}(a)$ in dB, compared to the SNR of an equivalent active RIS, i.e., in this standard operational regime, its SNR increases with the number of active elements, but it cannot reach that of active RIS, since this would require $a=1 \Leftrightarrow N_1=N$ and $N_2=0$, that is, an active RIS. Altering the fraction of active elements in active/passive RIS from $a=a_1$ to $a=a_2>a_1$ results in a relative SNR loss of $10\log_{10}(a_1/a_2)$. For example, with $a=0.125$, the SNR loss is about 9 dB, while with $a=0.875$, it is about 0.58 dB, i.e., from $a=0.875$ to $a=0.125$, there is a relative SNR loss of about $9-0.58=8.42$ dB, as validated also in Fig.~\ref{subfig:2a}. Note that active/passive RIS reaches 99.2\% of active RIS's SNR with 12.5\%  of elements being passive.

Next, we consider an impractically large number of elements, $N=20\times 10^6$. We note in Fig.~\ref{subfig:2b} that the SNR contribution of passive RS is about 19.95 dB higher in this regime than that of active RS for $a=0.125$; thus, the approximation in Eq.~(\ref{eq:11}) does not hold. As the number of active elements in active/passive RIS increases, and at the same time the number of passive elements correspondingly decreases such that $N_1 + N_2 = N$, the SNR of the active RS increases linearly for $a\geq 0.25$ due to Eq.~(\ref{eq:7}), but the SNR of the passive RS drops quadratically for $a\leq 0.75$ due to Eq.~(\ref{eq:9}). Eventually, for $a=0.875$, the SNR of the passive RS is about 5.44 dB lower than that of the active RS. The two are equal at around $a=0.78$. We also notice that at $a=0.125$, the SNR of the passive RS is so large, due to the immense number of passive elements which results in large reflect beamforming gain, that active/passive RIS outperforms active RIS, whose SNR is hit by amplification noise, by about 10.92 dB. Since as we increase $a$, the SNR of passive RS drops faster than that of active RS grows, the SNR of active/passive RIS drops with the number of active elements in this regime. It is only about 0.5 dB higher than  that of active RIS at $a=0.875$. 
 
\textbf{Remark 2:} By letting $P_{r,1}^{\max}\rightarrow\infty$, the asymptotic SNR of active/passive RIS is upper-bounded by\footnote{Similar to~\cite{IRS1,IRSAct8,Subconnected}, this implies in practice setting the reflect power budget to a large, but within the linear operation regime of the power amplifiers, value. This applies to the transmit power budget as well. The SNR impact of amplifiers' non-linearities is an interesting future research topic.} 
\begin{equation}\label{eq:add2}
\gamma_{a/p}^{\infty_{r}}\left(N,\frac{P_{\max}}{2}\right) \rightarrow\gamma_{a}^{\infty_{r}}\left(aN,\frac{P_{\max}}{2}\right)=a\gamma_{a}^{\infty_{r}}\left(N,\frac{P_{\max}}{2}\right),
\end{equation}
similar to the standard operation regime, where 
\begin{equation}\label{eq:12}
\gamma_{a}^{\infty_{r}}\left(N,\frac{P_{\max}}{2}\right) = N \frac{P_{\max}\pi^2 \varrho_g^2}{32 \delta^2}.
\end{equation}
For large values of $P_{t,a/p}^{\max}$, in turn, we have 
\begin{align}\label{eq:add3}
\gamma_{a/p}\left(N,\frac{P_{\max}}{2},\frac{P_{\max}}{2}\right)&\approx\gamma_{p}\left((1-a)N,\frac{P_{\max}}{2}\right)\nonumber \\
&=(1-a)^{2}\gamma_{p}\left(N,\frac{P_{\max}}{2}\right),
\end{align}
i.e., as we increase $P_{t,a/p}^{\max}$, active RS's impact is reduced  until it becomes negligible while passive RS's is pronounced. Then, active/passive RIS presents an SNR loss of $0.5*(1-a)^{2}$, or $20\log_{10}(1-a)+3$ in dB, compared to passive RIS. The extra $10\log_{10}(0.5)=3$ dB of SNR loss are attributed to the fact that the transmit power budget is $P_{\max}/2$, while in the passive RIS setup it is $P_{\max}$. For instance, when $a=0.125$, that is, $a-1\triangleq b = 0.875$, the SNR loss is about 4.16 dB, while for $b=0.125$, it is about 21 dB, i.e., the relative SNR loss attributed to the change from $b=b_1=0.875$ to $b=b_2=0.125$, $20\log_{10}\left(b_2/b_1\right)$, is about $21-4.16=16.84$ dB, as validated by the simulation results in Fig.~\ref{subfig:2c} for $N=256$. That is, the SNR of active/passive RIS increases in this regime  with the number of elements in passive RS, but it never reaches that of its passive RIS counterpart, since this would require $a=0 \Leftrightarrow N_2 = N$ and $N_1=0$, i.e., a passive RIS, as well as allocating all radiated power budget to the BS

\textbf{Lemma 2 (Asymptotic SNR for active/active RIS):} The SNR for arbitrary number of partitions $S$ is given by
\begin{align}\label{eq:18}
    &\gamma_{a/a}\left(N,P_{t,a/a}^{\max},P_{r,s}^{\max}\right) \rightarrow \nonumber \\
    &\sum_{s\in\mathcal{S}}N_s\frac{P_{t,a/a}^{\max}P_{r,s}^{\max}\pi^{2}\varrho_{f_s}^{2}\varrho_{g_s}^{2}}{16\left(P_{r,s}^{\max}\delta_s^{2}\varrho_{f_s}^{2}+4P_{t,a/a}^{\max}\sigma^{2}\varrho_{g_s}^{2}+4\sigma^{2}\delta_s^{2}\right)} \nonumber \\
&=\frac{N}{S}\sum_{s\in\mathcal{S}}\frac{P_{\max}^2 \pi^2 \rho_{f_s}^2 \rho_{g_s}^2}{32\left[P_{\max}\delta_s^2\rho_{f_s}^2 + 4S\left(P_{\max}\sigma^2\rho_{g_s}^2 + 2\sigma^2\delta_s^2\right)\right]}.
\end{align}

\textit{Proof:} Please see Appendix~\ref{App:B}. \qed

\textbf{Remark 3:} We note that $\gamma_{a/a}\left(N,P_{\max}/2,P_{\max}/2S\right) \rightarrow (1/S)\sum_{s\in\mathcal{S}}\gamma_{a}^{(s)}\left(N,P_{\max}/2,P_{\max}/2S\right)$,
i.e., the SNR of active/active RIS is the average of the SNR of $S$ active RISs with $N$ elements and reflect power budget $P_r^{\max}/S$ each. If $\varrho_{f_s}^2 = \varrho_f^2$, $\varrho_{g_s}^2 = \varrho_g^2$, and $\delta_s^2=\delta^2$, $\forall s\in\mathcal{S}$, then $\gamma_{a}^{(s)} = \gamma_{a}$. Thus,
\begin{figure*}[!t]
\centering
\subfloat[]{
	\label{subfig:3a}
	\includegraphics[scale = 0.3]{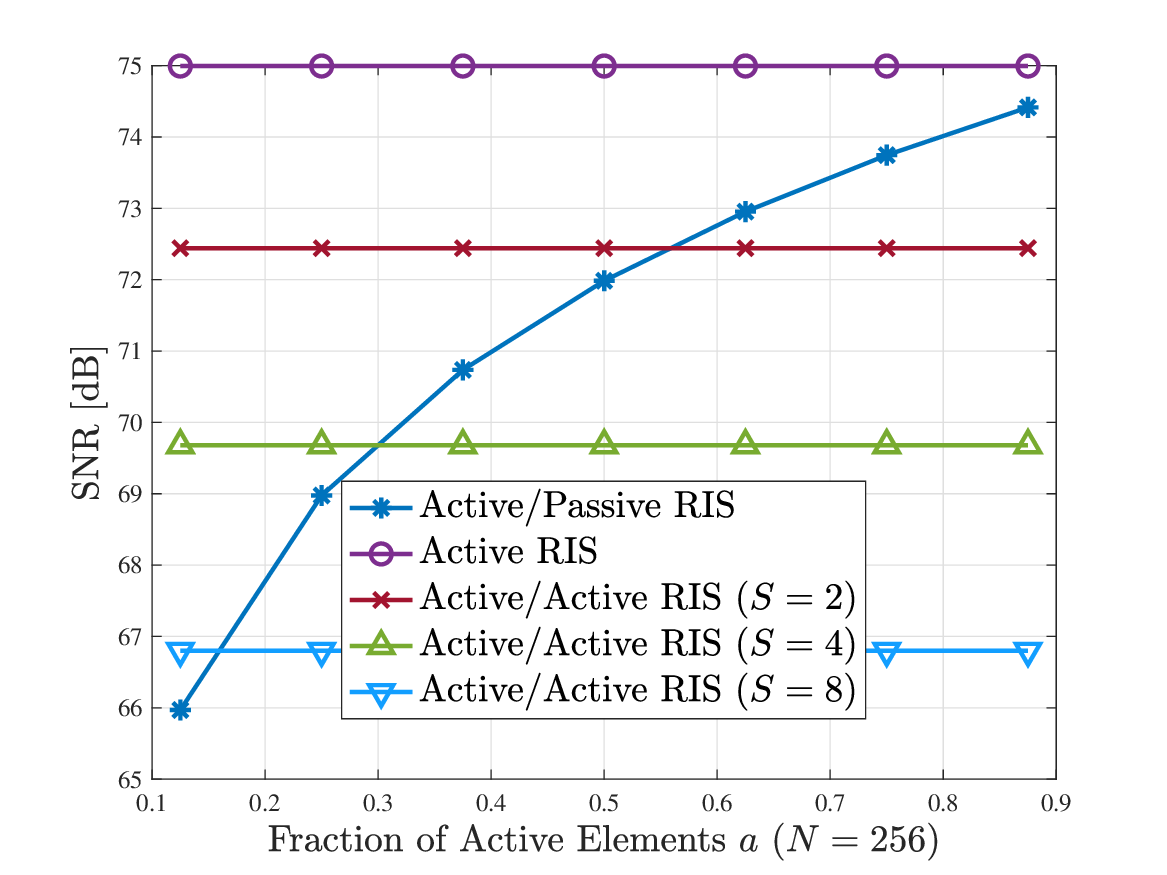} } 
\subfloat[]{
	\label{subfig:3b}
	\includegraphics[scale = 0.3]{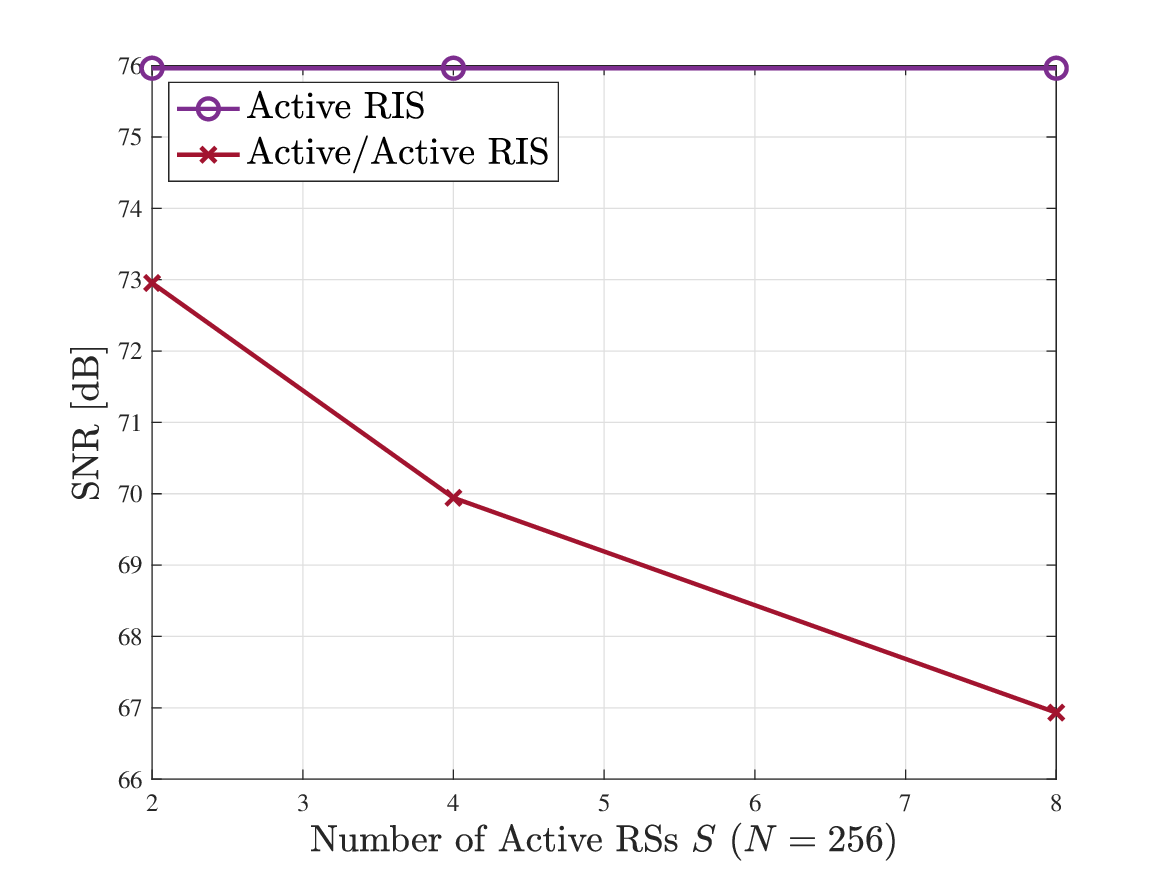} }
\subfloat[]{
	\label{subfig:3c}
	\includegraphics[scale = 0.3]{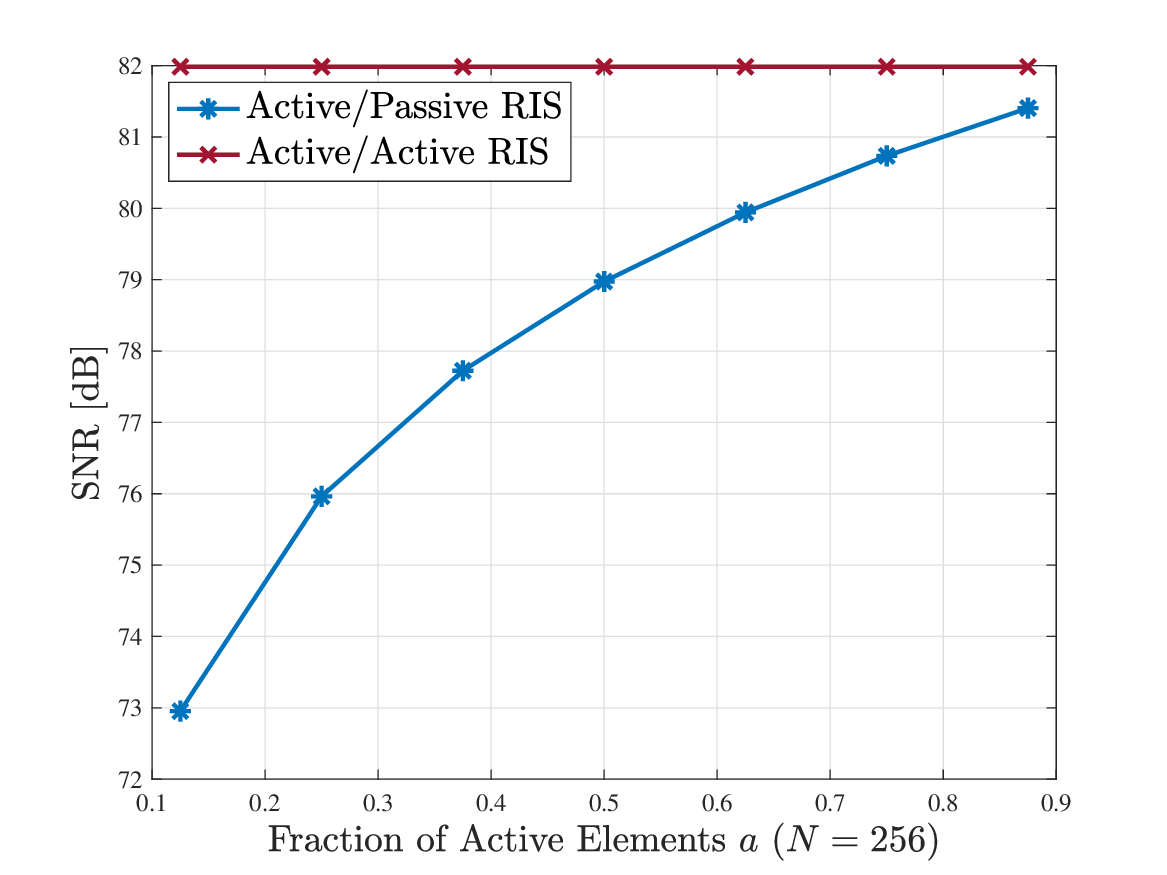} }  
\caption{SNR of active/active vs. active or active/passive RIS as we vary the number of RSs in active/active RIS or the fraction of active elements in active/passive RIS, for different operation regimes: (a) standard regime, (b) large transmit power budget, and (c) large reflect power budget.}
\label{fig:3}
\end{figure*}
\begin{align*}
&\gamma_{a/a}\left(N,\frac{P_{\max}}{2},\frac{P_{\max}}{2S}\right) \rightarrow \\
&\frac{1}{S}\left(S\gamma_{a}\left(N,\frac{P_{\max}}{2},\frac{P_{\max}/2}{S}\right)\right) = \gamma_{a}\left(N,\frac{P_{\max}}{2},\frac{P_{\max}/2}{S}\right)
\end{align*}
\begin{equation}\label{eq:20}
= N\frac{P_{\max}^2 \pi^2 \rho_{f}^2 \rho_{g}^2}{32\left[P_{\max}\delta^2\rho_{f}^2 + 4S\left(P_{\max}\sigma^2\rho_{g}^2 + 2\sigma^2\delta^2\right)\right]}.
\end{equation}
That is, the SNR of an active/active RIS with $N$ elements and $S$ active RSs equals that of a fully-active RIS with $N$ elements and a reflect power budget of $P_{r}^{\max}/S$ instead of $P_{r}^{\max}$, i.e., $S$ times smaller reflect power budget than usual. Consequently, active/active RIS presents an SNR loss in comparison to active RIS. Its SNR  improves by reducing the number of active RSs, but it never reaches that of active RIS, since this would require $S=1$, i.e., an active RIS (a hybrid RIS implies $S\geq 2$). Nevertheless, it is non-trivial to quantify the aforementioned SNR loss, due to the fact that $P_{r,s}^{\max}$ is involved both at the nominator and at one term of the denominator in the SNR expression, as indicated by Eq.~(\ref{eq:18}). Thus, we will rely instead on numerical simulations. As shown in Fig.~\ref{subfig:3a}, the SNR loss of an active/active RIS with $N=256$ elements and $S=2$, $S=4$, or $S=8$ active RSs, compared to the SNR of an equivalent active RIS, is 2.55 dB, 5.32 dB, or 8.2 dB, respectively. That is, active/active RIS reaches up to about 96.6\% of active RIS's SNR, as a result of the reduction in the reflect power budget per RS when going from $S=1$ (active RIS) to $S=2$ (active/active RIS). We also note that active/passive RIS reaches the SNR of active/active RIS with $S=2$, $S=4$, or $S=8$ active RSs when the fraction of active elements is around $a=0.55$, $a=0.3$, and $a=0.15$, respectively. Thus, active/passive RIS with a single power amplifier needs 141 active elements to reach the SNR of active/active RIS with 2 power amplifiers in total, showing slow SNR growth with the number of active elements. 

\textbf{Remark 4:} By letting $P_{t,a/a}^{\max}\rightarrow\infty$ in Eq.~(\ref{eq:18}), the asymptotic SNR of active/active RIS is upper-bounded by
\begin{equation}\label{eq:20New}
\gamma_{a/a}^{\infty_{t}}\left(N,P_{r,s}^{\max}\right)\rightarrow \frac{1}{S}\sum_{s\in\mathcal{S}}\frac{1}{S}N\frac{P_r^{\max}\pi^2\varrho_{f_s}^2}{64\sigma^2},
\end{equation}
where  we used the fact that $P_{r,s}^{\max}=P_r^{\max}/S$. Assuming that $\varrho_{f_s}^2=\varrho_f^2$, $\forall s\in\mathcal{S}$, we obtain
\begin{align}\label{eq:20a}
&\gamma_{a/a}^{\infty_{t}}\left(N,\frac{P_r^{\max}}{S}\right)\rightarrow \frac{1}{S}\left(S\cdot\frac{1}{S}\cdot N\frac{P_r^{\max}\pi^2\varrho_{f_s}^2}{64\sigma^2}\right) \nonumber \\
&=\frac{1}{S}N\frac{P_r^{\max}\pi^2\varrho_{f}^2}{64\sigma^2}=\frac{1}{S}\gamma_{a}^{\infty_{t}}\left(N,P_{r}^{\max}\right),
\end{align}
where $\gamma_{a}^{\infty_{t}}$ is the asymptotic SNR of an active RIS with $N$ elements and reflect power budget $P_r^{\max}$ in the large transmit power regime. Hence, in this regime, active/active RIS presents an SNR loss of $10\log_{10}(S)$ dB compared to active RIS, i.e., its SNR improves as we reduce the number of active RSs, but it never reaches that of active RIS, since this would require $S=1$, that is, an active RIS. The relative SNR loss attributed to the increase of the number of active RSs from $S=S_1$ to $S=S_2$ equals $10\log_{10}(S_2/S_1)$ dB. For instance, an active/active RIS with $N=256$ elements and $S=2$, $S=4$, or $S=8$ active RSs presents an SNR loss of 3.01 dB, 6.02 dB, or 9.03 dB, respectively, compared to an equivalent active RIS, i.e., each successive doubling in the number of active RSs results in a relative SNR loss of 3.01 dB, as shown in Fig.~\ref{subfig:3b}. We do not compare active/active RIS with active/passive RIS in Fig.~\ref{subfig:3b}, since in this regime the latter performs worse than passive RIS, as illustrated in Fig.~\ref{subfig:2c}.

Likewise, by letting $P_{r,s}^{\max}\rightarrow\infty$ in Eq.~(\ref{eq:18}), the asymptotic SNR of active/active RIS is upper-bounded by
\begin{equation}\label{eq:add5New}
\gamma_{a/a}^{\infty_{r}}\left(N,\frac{P_{\max}}{2}\right)\rightarrow
\frac{1}{S}\sum_{s\in\mathcal{S}}N \frac{P_{\max}\pi^2 \varrho_{g_s}^2}{32 \delta_s^2}.
\end{equation}
Assuming that $\varrho_{g_s}^{2}=\varrho_g^2$ and $\delta_s^2=\delta^2$, $\forall s\in\mathcal{S}$, we have:
\begin{align}\label{eq:add5}
&\gamma_{a/a}^{\infty_{r}}\left(N,\frac{P_{\max}}{2}\right) \rightarrow \frac{1}{S}\left(S\cdot N \frac{P_{\max}\pi^2 \varrho_g^2}{32 \delta^2}\right) \nonumber \\
&=N \frac{P_{\max}\pi^2 \varrho_g^2}{32 \delta^2}=\gamma_{a}^{\infty_{r}}\left(N,\frac{P_{\max}}{2}\right).
\end{align}
That is, for large reflect power budget, active/active RIS achieves the same SNR with active RIS, regardless of the number of active RSs. Therefore, it presents an SNR gain of $10\log_{10}(a)$ over active/passive RIS with a fraction $a$ of active elements in this regime, as shown by Eq.~(\ref{eq:add2}) and Fig.~\ref{subfig:3c}. 

\subsection{Impact of RIS Size}\label{subsec:3.2}
\textbf{Lemma 3 (RIS size---passive vs. active/passive RIS):} Taking into account power amplifier sharing and solving $\gamma_p \geq \gamma_a$ for $N$, we note that passive RIS outperforms active RIS in terms of SNR when
\begin{equation}\label{eq:21}
    N \geq \frac{P_{t,a}^{\max}}{P_{t,p}^{\max}}\frac{P_{r}^{\max}\sigma^{2}}{P_{r}^{\max}\delta^{2}\varrho_{f}^{2}+4P_{t,a}^{\max}\sigma^{2}\varrho_{g}^{2}+4\sigma^{2}\delta^{2}}\triangleq N_{\min}.
\end{equation}
Hence, based on the approximation in Eq.~(\ref{eq:11}), which is tight as shown in Fig.~\ref{subfig:2a}, passive RIS outperforms active/passive RIS for large $N$ and arbitrary $a$ when $N\geq aN_{\min}$.
\begin{figure*}[!t]
\centering
\subfloat[$N=0.25:0.25:1$ ($\times 10^{6}$).]{
	\label{subfig:4a}
	\includegraphics[scale = 0.3]{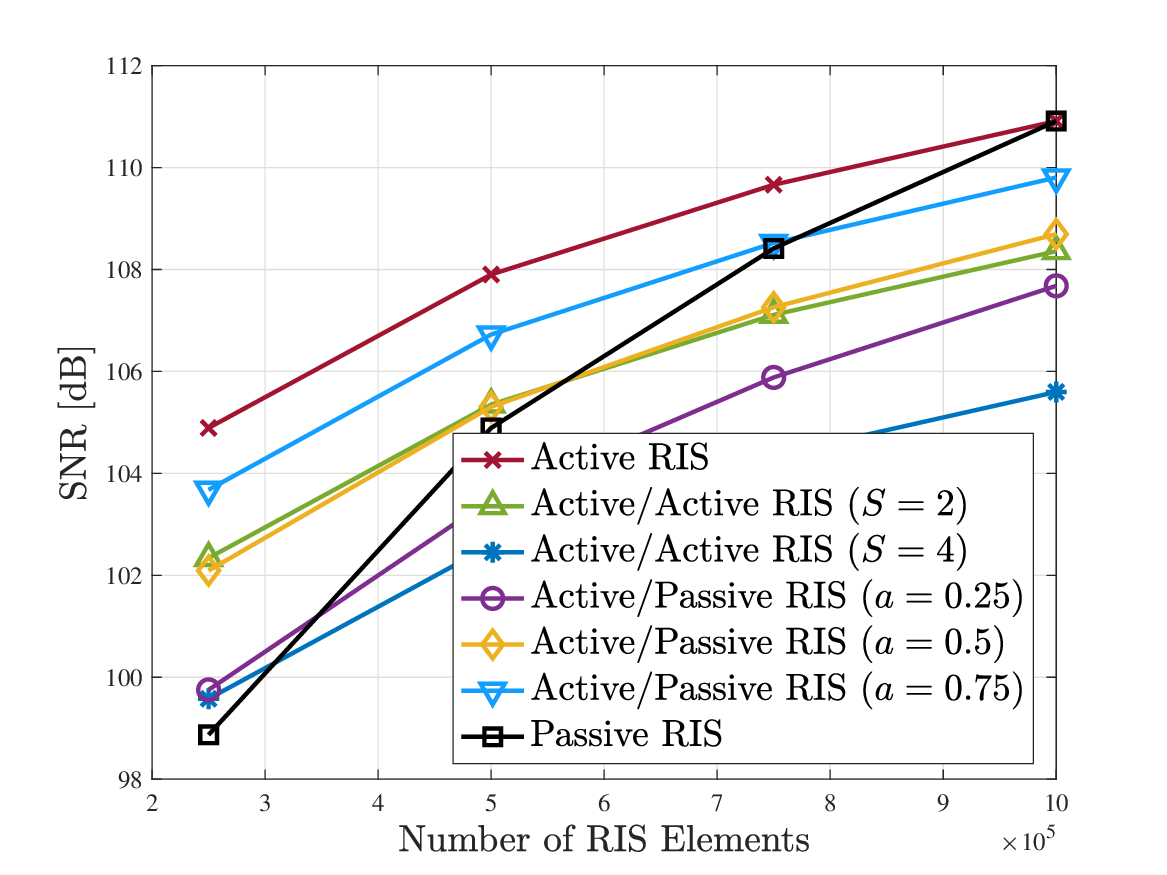} } 
 \subfloat[$N=2.5:2.5:10$ ($\times 10^{6}$).]{
	\label{subfig:4b}
	\includegraphics[scale = 0.3]{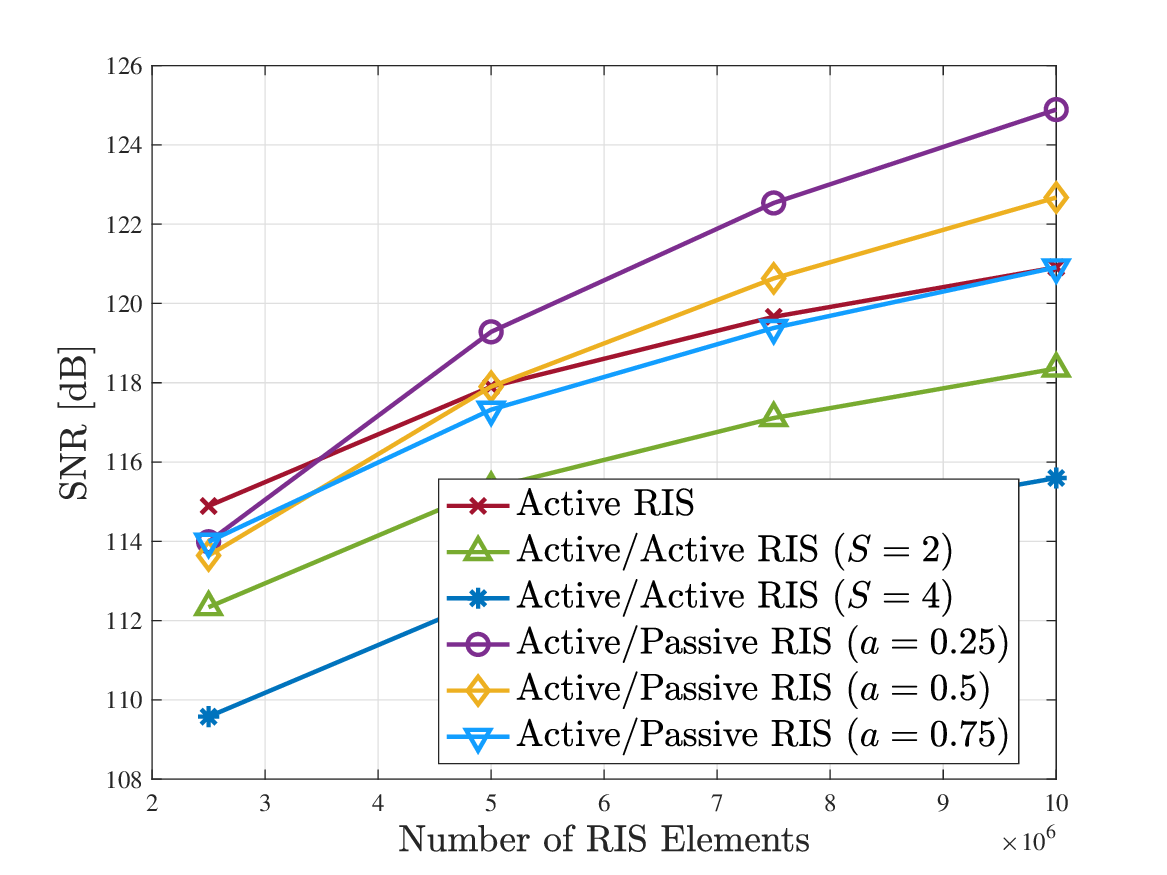} }
 \subfloat[$N=64:64:1,920$.]{
	\label{subfig:4c}
	\includegraphics[scale = 0.3]{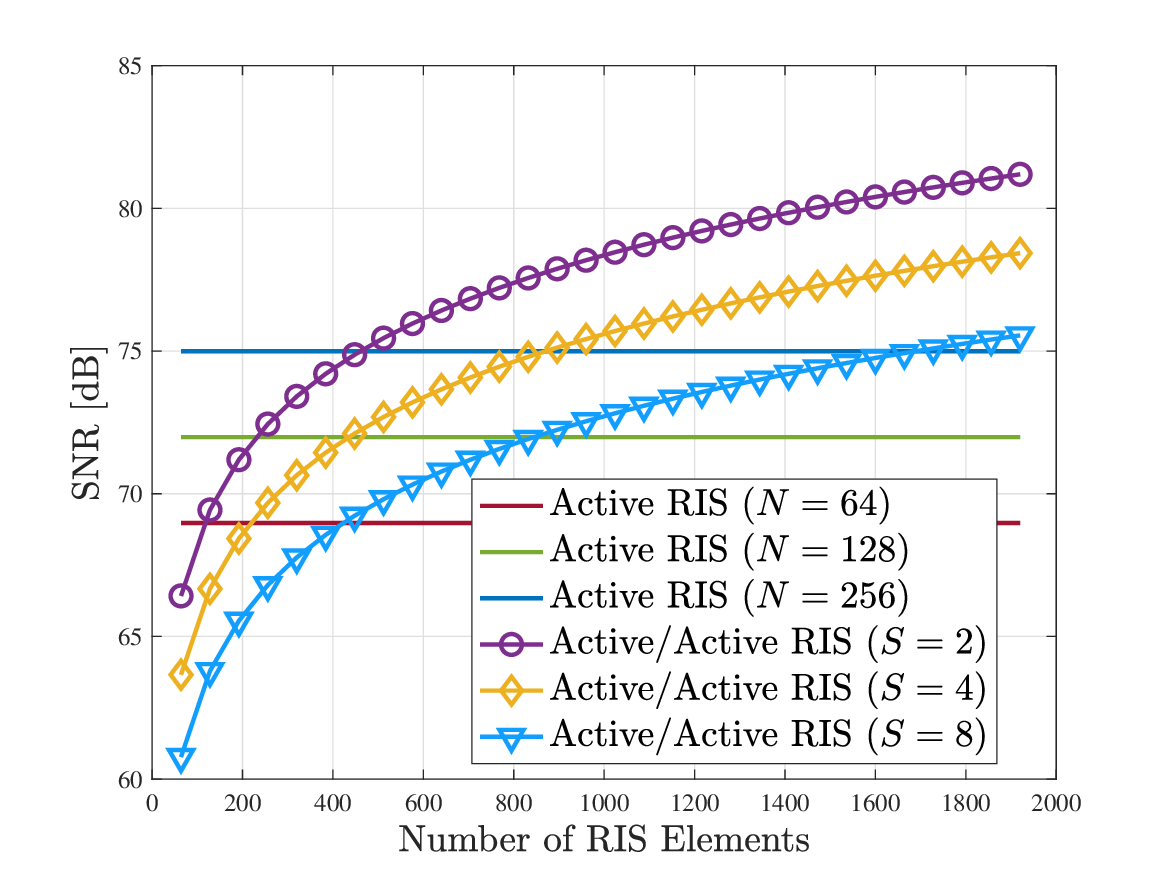} } 
\caption{Asymptotic SNR for the considered RIS architectures as the number of RIS elements varies.} 
\label{fig:4}
\end{figure*}

\textbf{Remark 5:} According to the running numeric example, passive RIS outperforms active RIS when $N \geq 1\times 10^6$. Consequently, it outperforms active/passive RIS with $a=0.25$, $a=0.5$, or $a=0.75$ when $N \geq 0.25\times 10^6$, $N \geq 0.5\times 10^6$, or $N \geq 0.75\times 10^6$, respectively, as shown in Fig.~\ref{subfig:4a}. Thus, fewer RIS elements are required, compared to the active RIS case, and this figure is reduced as the number of active elements in active/passive RIS is decreased, as expected.
\begin{figure*}[!t]
\normalsize
\setcounter{mytempeqncnt}{\value{equation}}
\setcounter{equation}{19}
\begin{equation}\label{eq:add8}
    N\geq \frac{P_{t,a}^{\max}P_{r}^{\max}\varrho_{f}^{2}\varrho_{g}^{2}\left(P_{r,1}^{\max}\delta_{1}^{2}\varrho_{f_{1}}^{2}+4P_{t,a/p}^{\max}\sigma^{2}\varrho_{g_{1}}^{2}+4\sigma^{2}\delta_{1}^{2}\right)-aP_{t,a/p}^{\max}P_{r,1}^{\max}\varrho_{f_{1}}^{2}\varrho_{g_{1}}^{2}\left(P_{r}^{\max}\delta^{2}\varrho_{f}^{2}+4P_{t,a}^{\max}\sigma^{2}\varrho_{g}^{2}+4\sigma^{2}\delta^{2}\right)}{4\left(1-a\right)^{2}P_{t,a/p}^{\max}\varrho_{f_{2}}^{2}\varrho_{g_{2}}^{2}\left(P_{t,a/p}^{\max}\varrho_{g_{1}}^{2}+\delta_{1}^{2}\right)\left(P_{r}^{\max}\delta^{2}\varrho_{f}^{2}+4P_{t,a}^{\max}\sigma^{2}\varrho_{g}^{2}+4\sigma^{2}\delta^{2}\right)}.
\end{equation}
\setcounter{equation}{\value{mytempeqncnt}}
\hrulefill
\vspace*{4pt}
\end{figure*}

\textbf{Lemma 4 (RIS size---passive vs. active/active RIS):} Based on Eq.~(\ref{eq:20}), passive RIS outperforms active/active RIS, in terms of SNR, for large $N$, arbitrary $S$, and $N_s=N/S$ when
\begin{equation}
N \geq \frac{\sigma^2 P_{\max}}{2P_{\max}\delta^2\rho_f^2 + 8S\sigma^2\left(P_{\max}\rho_g^2+2\delta^2\right)}.
\end{equation} 

\textit{Proof:} Solve $\gamma_p \geq \gamma_{a/a}$ for $N$. \qed

\textbf{Remark 6:} Passive RIS outperforms active/active RIS with $S=2$ or $S=4$ RSs when $N \geq 0.55\times 10^6$ or $N \geq 0.29\times 10^6$, respectively, as shown in Fig.~\ref{subfig:4a}. The reduction in the required number of elements is due to the negative impact on the SNR of the smaller reflect power in each RS for higher number of RSs. We note in Fig.~\ref{subfig:4a} that active/active RIS with $S=2$ or $S=4$ slightly outperforms active/passive RIS with $a=0.5$ or $a=0.25$, respectively. 

\textbf{Lemma 5 (RIS size---active/passive vs. active RIS):} Based on Eq.~(\ref{eq:5}), active/passive RIS outperforms active RIS, in terms of SNR, for large $N$ and arbitrary $a$ when the number of elements satisfies Eq.~(\ref{eq:add8}) at the top of this page.

\textit{Proof:} Solve $\gamma_{a/p} \geq \gamma_{a}$ for $N$. \qed

\textbf{Remark 7:} We note in Fig.~\ref{subfig:4b} that active/passive RIS with $a=0.75$, $a=0.5$, or $a=0.25$ outperforms active RIS when $N \geq 10\times 10^6$, $N \geq 5\times 10^6$, or $N \geq 3.33\times 10^6$, respectively, i.e., the more active elements, the more elements in total are required! Furthermore, while active/passive RIS substantially outperforms passive RIS, being only $10\log_{10}(a)$ dB away from the SNR of active RIS, it needs a much larger number of elements than passive RIS to reach that SNR! For large $N$ and especially under $N_2\gg N_1$, the significant SNR contribution of passive RS can be expressed as follows:
\addtocounter{equation}{1}
\begin{equation}
c\gamma_{p}\left(N_{2},P_{t,a/p}^{\max}\right) =c^{\prime}N_2\gamma_a\left(N_{2},P_{t,a/p}^{\max}\right),
\end{equation}
where $c^{\prime}=4\left(\varrho_{g_1}^2P_{t,a/p}^{\max}+\delta_1^2\right)/P_{r,1}^{\max}$. Thus, the SNR contribution of the passive RS equals that of an active RIS with $N_{2}$ elements times $c^{\prime} N_2$, i.e., it is directly proportional to $N_{2}$. This explains why a smaller total number of elements is needed for active/passive RIS to reach the SNR of active RIS when $N_2\gg N_1$. Nevertheless, since $N_{2}<N$ by definition, and $c^{\prime}$ is extremely small (e.g., $\approx 4\times 10^{-7}$ in our numeric example), still a much larger number of elements than in the passive RIS case is required to reach active RIS performance.

\textbf{Lemma 6 (RIS size---active/active vs. active RIS):} Since $P_{t,a/a}^{\max}=P_{t,a}^{\max}$, active/active RIS with $N$ elements and $S$ RSs outperforms active RIS with $N_{a}$ elements when
\begin{equation}\label{eq:25}
    N \geq N_{a}\frac{P_{r}^{\max}\delta^{2}\varrho_{f}^{2}+4S\sigma^{2}\left(P_{t,a}^{\max}\varrho_{g}^{2}+\delta^{2}\right)}{P_{r}^{\max}\delta^{2}\varrho_{f}^{2}+4\sigma^{2}\left(P_{t,a}^{\max}\varrho_{g}^{2}+\delta^{2}\right)}.
\end{equation}

\textit{Proof:} Set $\gamma_{a/a}\geq\gamma_{a}$, i.e., $\gamma_{a}\left(N,P_{t,a}^{\max},P_{r}^{\max}/S\right)\geq\gamma_{a}\left(N_{a},P_{t,a}^{\max},P_{r}^{\max}\right)$, and solve for $N$. \qed

\textbf{Remark 8:} Due to the smaller reflect power budget of each RS, active/active RIS reaches the SNR of an active RIS with \textit{the same total radiation power budget} $P_{\max}$ only if it has additional elements. More elements are required for higher number of RSs. Specifically, if $S_i=2S_{i-1}$, then $N_i = N_{i-1}+i\cdot 1.6N_a$, $i=1,2,\dots$, e.g., for $\left\{S_0,S_1,S_2\right\}=\{2,4,8\}$, $N=\{1.8,3.4,6.6\}\times N_a$, as shown in Fig.~\ref{subfig:4c}.

\subsection{Asymptotic SNR in LoS Scenario}\label{subsec:3.4}
We have considered so far Rayleigh fading, which provides mathematical tractability. Here, we present the asymptotic SNR results for the other extreme scenario of Line-of-Sight (LoS) channels. The derivations follow closely the given proofs and are omitted.

\textbf{Lemma 7 (Asymptotic SNR for active/passive RIS in LoS Scenario):} The SNR for arbitrary elements' allocation, assuming LoS channels, is given by
\begin{align}\label{eq:LoS1}
\gamma_{a/p} \rightarrow & aN\frac{P_{\max}^2\varrho_{f_1}^2\varrho_{g_1}^2}{4\left(P_{\max}\delta_1^2\varrho_{f_1}^2
+ P_{\max}\sigma^2\varrho_{g_1}^2 + 2\sigma^2\delta_1^2\right)} \nonumber \\
&+ (1-a)^2N^2\frac{P_{\max}\varrho_{f_2}^2\varrho_{g_2}^2}{8\sigma^2}.
\end{align}

\textbf{Lemma 8 (Asymptotic SNR for active/active RIS in LoS Scenario):} The SNR for arbitrary number of partitions $S$, assuming LoS channels, is given by
\begin{equation}\label{eq:LoS2}
\gamma_{a/a} \rightarrow \frac{N}{S}\sum_{s\in\mathcal{S}}\frac{P_{\max}^2\varrho_{f_s}^2\varrho_{g_s}^2}{4\left(P_{\max}\delta_s^2\varrho_{f_s}^2 + SP_{\max}\sigma^2\varrho_{g_s}^2 + 2S\sigma^2\delta_s^2\right)}.
\end{equation}

\textbf{Remark 9:} We note that the factor $\pi^2/16$ is absent in the LoS SNR expressions, since there's no averaging over random channel realizations. The SNR scaling with $N$ remains the same, linear for active RS and quadratic for passive. 

\subsection{Conclusions}\label{subsec:3.3}
The asymptotic analysis offers key insights into the performance of hybrid RIS architectures with a single reflect-type power amplifier per RS, compared to passive RIS and single-amplifier active RIS. The latter achieves higher asymptotic SNR, with active/passive RIS requiring an impractical number of elements to match it and active/active RIS with $S=2$ RSs needing about 1.8 times more elements than active RIS under the same total radiated power budget. Thus, active RIS is preferable in SNR-focused scenarios. On the other hand, both hybrid designs achieve over 96\% of active RIS’s performance. Hence, considering the TPC savings of these hybrid RIS structures compared to RIS designs with a power amplifier per active element and the diminishing returns caused by increasing amplifiers count, significant EE gains are projected for the aforementioned SC-active RS-based hybrid architectures, making them suitable when EE is the priority. Specifically, we expect hybrid SC-active RIS to offer the best EE, balancing performance and power consumption; hybrid SC-active/passive RIS to be preferred under strict TPC constraints, as it outperforms passive RIS while maintaining low power usage; and FC-active/SC-active RIS to suit performance-focused scenarios where power savings are required. However, a comprehensive study of EE and capacity performance is needed to confirm these projections, thus leading us to the next section.

\section{Problem Formulation and Solutions}\label{sec:4}
In this section, we are interested in joint precoding/RIS beamforming optimization, such that the EE of a hybrid RIS-aided multi-user MISO downlink system is maximized subject to the BS's and RIS's TPC constraints. This is a non-trivial task. The novel mathematical representation of power amplifier sharing and the system model introduced in Sec.~\ref{sec:2}, where everything is described in terms of each RS's beamforming matrix instead of the composite block-diagonal RIS beamforming matrix, has driven and simplified optimization, essentially allowing us to treat each RS as an individual RIS itself and apply well-known optimization methods. This is a great advantage of the proposed approach, from both a practical and an insights perspective.

\subsection{Hybrid Active/Passive RIS}\label{subsec:4.1}
In the hybrid active/passive RIS case, the optimization problem of interest is mathematically formulated as follows:
\begin{subequations}\label{eq:26}
\begin{alignat}{2}
&&&\text{(P1): }\underset{\mathbf{w},\left\{\bm{\phi}_{s}\right\}}{\max} \ \eta=\frac{R}{P}=\frac{\sum\limits_{k\in\mathcal{K}}\log_{2}\left(1+\gamma_{k}\right)}{P_{\text{BS}}+P_{1}+N_{2}P_{\text{PS}}} \label{eq:26a} \\
&&&\text{s.t.} \ \ \ \ \text{C1: }P_{\text{BS}} = \xi^{-1}\sum_{k\in\mathcal{K}}\left\|\mathbf{w}_{k}\right\|^{2} + W_{\text{BS}} \leq P_{\text{BS}}^{\max}, \label{eq:26b} \\
&&& \ \ \ \ \ \ \ \text{C2: }P_{1} = \zeta_{1}^{-1}\left(\sum_{k\in\mathcal{K}}\left\|\bm{\Phi}_{1}\mathbf{G}_{1}\mathbf{w}_{k}\right\|^{2} + \delta_{1}^{2}\left\|\bm{\Phi}_{1}\right\|_{F}^{2}\right) \nonumber \\
&&& \ \ \ \ \ \ \ \ \ \ \ \ \ \ +W_{r,1} \leq P_{1}^{\max}, \label{eq:26c} \\
&&& \ \ \ \ \ \ \ \text{C3: }\left|\left[\bm{\phi}_{2}\right]_{n}\right|=1, \ \forall n\in\mathcal{N}_{2}, \label{eq:26d}
\end{alignat}
\end{subequations}
where we have defined $\mathbf{w}\triangleq\left[\mathbf{w}_{1}^{T},\dots,\mathbf{w}_{K}^{T}\right]^{T}\in\mathbb{C}^{KM}$ and $\bm{\phi}_{s}\triangleq\operatorname{Diag}\left(\bm{\Phi}_{s}^{*}\right)\in\mathbb{C}^{N_{s}}$, $\gamma_{k}$ is given by Eq.~(\ref{eq:3}), and $W_{r,1}=N_{1}\left(P_{\text{PS}}+P_{\text{DC}}\right)$ (FC) or $ W_{r,1}=N_{1}P_{\text{PS}}+L_1 P_{\text{DC}}$ (SC). (P1) is a challenging non-convex optimization problem, due to the fractional form of the objective function (OF), the intrinsic coupling of the decision variables in the OF and C2, and the unit modulus constraints (UMC) C3.

\subsubsection{Fractional Programming}\label{subsubsec:4.1.1}
Using Dinkelbach's algorithm, we transform the OF into $f\left(\mathbf{w},\bm{\phi}\right) = R-\eta P$, where $\bm{\phi}\triangleq\left[\bm{\phi}_{1}^{T},\bm{\phi}_{2}^{T}\right]^{T}$~\cite{Subconnected}. Then, by applying the Lagrangian dual transform and the quadratic transform~\cite{FP1}, we recast (P1) as 
\begin{equation}\label{eq:27}
    \text{(P2): }\underset{\mathbf{w},\bm{\phi},\bm{\mu},\bm{\nu}}{\max} \ g\left(\mathbf{w},\bm{\phi},\bm{\mu},\bm{\nu}\right) \ \text{s.t. C1--C3},
\end{equation}
where $\bm{\mu}\in\mathbb{C}^{K}$ and $\bm{\nu}\in\mathbb{C}^{K}$ are auxiliary variables, $g\left(\mathbf{w}, \bm{\phi}, \bm{\mu}, \bm{\nu}\right) = -\eta P+\sum_{k\in\mathcal{K}}\ln \left(\widebar{\mu}_k\right)-\mu_k +2 \sqrt{\widebar{\mu}_k}\rho_k-\sum_{k\in\mathcal{K}}\left|\nu_k\right|^2\left(\sum_{i\in\mathcal{K}}\left|\mathbf{h}_{k}^{\dagger}\mathbf{w}_i\right|^2+\left\|\mathbf{f}_{k,1}^{\dagger} \bm{\Phi}_{1}\right\|^2 \delta_1^2+\sigma_{k}^2\right)$, $\widebar{\mu}_k\triangleq1+\mu_k$, and $\rho_{k}\triangleq\operatorname{Re}\left\{\nu_{k}^{*}\mathbf{h}_{k}^{\dagger}\mathbf{w}_{k}\right\}$. Next, we apply the BCA method to alternately update each variable with the other variables fixed, as described below. The BCA method has been selected because it is simple to implement, well-studied, and allows for direct comparisons with other works and RIS architectures. Furthermore, numerical simulations show that we obtain high-quality solutions across a variety of test scenarios after a few iterations, as we shall see in Sec.~\ref{sec:5}.

\subsubsection{Blocks Update}\label{subsubsec:4.1.2}
\paragraph{Update Auxiliary Variables}
Solving $\partial g/\partial \mu_{k}=0$ and $\partial g/\partial \nu_{k}=0$ with fixed $\left(\mathbf{w},\bm{\phi},\bm{\nu}\right)$ and $\left(\mathbf{w},\bm{\phi},\bm{\mu}\right)$, respectively, we obtain:
\begin{subequations}\label{eq:29}
\begin{alignat}{2}
\mu_k^{\star} & =\frac{\rho_{k}}{2}\left(\rho_k+\sqrt{\rho_k^2+4}\right), \label{eq:29a} \\
\nu_k^{\star} & =\frac{\sqrt{\widebar{\mu}_k}\mathbf{h}_{k}^{\dagger} \mathbf{w}_k}{\sum\limits_{i\in\mathcal{K}}\left|\mathbf{h}_{k}^{\dagger} \mathbf{w}_i\right|^2+\left\|\mathbf{f}_{k,1}^{\dagger} \bm{\Phi}_{1}\right\|^2 \delta_1^2+\sigma_{k}^2}. \label{eq:29b}
\end{alignat}
\end{subequations}

\paragraph{Update Transmit Precoding}
From Eqs.~(\ref{eq:26b}) and~(\ref{eq:26c}), $\widetilde{P}_{\text{BS}}^{\max} \triangleq \xi\left(P_{\text{BS}}^{\max}-W_{\text{BS}}\right)$ and $\widetilde{P}_{1}^{\max} \triangleq \zeta_{1}\left(P_{1}^{\max}-W_{r,1}\right)$ are the transmit and reflect power budget of the BS and RS1, respectively. Let us respectively define $\mathbf{H}_{k}\in\mathbb{C}^{M\times M}$, $\mathbf{u}_{k}\in\mathbb{C}^{M}$, $\mathbf{u}\in\mathbb{C}^{KM}$, $\mathbf{E}_1\in\mathbb{C}^{N_1\times M}$, $\mathbf{T}_1\in\mathbb{C}^{KM\times KM}$, and $\mathbf{S}\in\mathbb{C}^{KM\times KM}$ as $\mathbf{H}_{k}\triangleq\mathbf{h}_{k}\mathbf{h}_{k}^{\dagger}$, $\mathbf{u}_{k}\triangleq 2\sqrt{\widebar{\mu}_k}\nu_{k}\mathbf{h}_{k}$, $\mathbf{u}\triangleq\left[\mathbf{u}_{1}^{T},\dots,\mathbf{u}_{K}^{T}\right]^{T}$, $\mathbf{E}_1\triangleq\bm{\Phi}_1\mathbf{G}_1$, $\mathbf{T}_1 \triangleq \mathbf{I}_{K}\otimes \left(\mathbf{E}_{1}^{\dagger}\mathbf{E}_{1}\right)$, and 
\begin{equation}\label{eq:31}
\mathbf{S} \triangleq \mathbf{I}_{K}\otimes\left(\eta\xi^{-1}\mathbf{I}_{M}+\eta\zeta_{1}^{-1}\mathbf{E}_{1}^{\dagger}\mathbf{E}_{1}+\sum_{k\in\mathcal{K}}\left|\nu_k\right|^2\mathbf{H}_{k}\right).
\end{equation}
Hence, with fixed $\left(\bm{\phi},\bm{\mu},\bm{\nu}\right)$, the transmit precoding optimization sub-problem can be formulated as follows:
\begin{subequations}\label{eq:32}
    \begin{alignat}{2}
        &&&\text{(P3): } \underset{\mathbf{w}}{\max} \ \operatorname{Re}\left\{\mathbf{u}^{\dagger}\mathbf{w}\right\}-\mathbf{w}^{\dagger}\mathbf{S}\mathbf{w} \label{eq:32a} \\
        &&&\text{s.t.} \ \ \ \ \text{C1: }\mathbf{w}^{\dagger}\mathbf{w} \leq \widetilde{P}_{\text{BS}}^{\max}, \label{eq:32b} \\
        &&& \ \ \ \ \ \ \ \text{C2: } \mathbf{w}^{\dagger}\mathbf{T}_1\mathbf{w} \leq \widetilde{P}_{1}^{\max} - \delta_{1}^{2}\left\|\bm{\Phi}_{1}\right\|_{F}^{2}. \label{eq:32c}
    \end{alignat}
\end{subequations}
(P3) is a standard convex Quadratically Constrained Quadratic Program (QCQP), which can be solved by using the Lagrange multipliers method  to obtain~\cite{IRSAct8}
\begin{equation}\label{eq:34}
\mathbf{w}^{\star} = \frac{1}{2}(\mathbf{S}+\lambda_{1}\mathbf{I}_{KM}+\lambda_{2}\mathbf{T}_1)^{-1}\mathbf{u},
\end{equation}
where the Lagrange multipliers corresponding to the constraints C1 and C2, $\lambda_{1}$ and $\lambda_{2}$, are optimized via grid search.

\paragraph{Update RIS Beamforming}
Let $\alpha_{k,i}\triangleq\mathbf{g}_{k}^{\dagger}\mathbf{w}_{i}$. We define $\bm{\pi}_{s,i}\in\mathbb{C}^{N_s}$ as $\bm{\pi}_{s,i}\triangleq\mathbf{G}_{s}\mathbf{w}_{i}$. Then, $\mathbf{h}_{k}^{\dagger}\mathbf{w}_{i}\triangleq\alpha_{k,i}+\sum_{s\in\mathcal{S}}\mathbf{f}_{k,s}^{\dagger}\operatorname{diag}\left(\bm{\pi}_{s,i}\right)\bm{\phi}_{s}$. We also define $\mathbf{F}_{k,s}\in\mathbb{C}^{N_s\times N_s}$, $\widebar{\mathbf{F}}_{k,1}\in\mathbb{C}^{N_1\times N_1}$, and $\widebar{\bm{\Pi}}_{1,k}\in\mathbb{C}^{N_1\times N_1}$ as $\mathbf{F}_{k,s}\triangleq\mathbf{f}_{k,s}\mathbf{f}_{k,s}^{\dagger}$, $\widebar{\mathbf{F}}_{k,1}\triangleq\operatorname{diag}\left(\mathbf{f}_{k, 1} \odot \mathbf{f}_{k, 1}^*\right)$, and $\widebar{\bm{\Pi}}_{1,k}\triangleq\operatorname{diag}\left(\boldsymbol{\pi}_{1, k} \odot \boldsymbol{\pi}_{1, k}^*\right)$. In addition, let us respectively define $\bm{\upsilon}_{s}\in\mathbb{C}^{N_{s}}$, $\mathbf{Q}_{s}\in\mathbb{C}^{N_{s}\times N_{s}}$, and $\mathbf{R}\in\mathbb{C}^{N_{1}\times N_{1}}$ as 
\begin{subequations}\label{eq:35}
    \begin{alignat}{2}
        \bm{\upsilon}_{s} \triangleq & \sum_{k\in\mathcal{K}}\operatorname{diag}\left(\mathbf{f}_{k,s}^{\dagger}\right)\left(2\sqrt{\widebar{\mu}_k}\nu_{k}^{*}\bm{\pi}_{s,k}-\left|\nu_{k}\right|^{2}\sum_{i\in\mathcal{K}}\alpha_{k,i}^{*}\bm{\pi}_{s,i}\right), \label{eq:35a} \\
        \mathbf{Q}_{1} \triangleq & \sum_{k\in\mathcal{K}}\left|\nu_{k}\right|^{2}\left(\delta_{1}^{2}\widebar{\mathbf{F}}_{k,1}+\sum_{i\in\mathcal{K}}\operatorname{diag}\left(\bm{\pi}_{1,i}^{*}\right)\mathbf{F}_{k,1}\operatorname{diag}\left(\bm{\pi}_{1,i}\right)\right)\nonumber \\
        &+\sum_{k\in\mathcal{K}}\eta\zeta_{1}\left(\widebar{\bm{\Pi}}_{1,k}+\delta_{1}^{2}\mathbf{I}_{N_{1}}\right), \label{eq:35b}
    \end{alignat}
\end{subequations}
$\mathbf{Q}_{2} \triangleq \sum_{k\in\mathcal{K}}\left|\nu_{k}\right|^{2}\sum_{i\in\mathcal{K}}\operatorname{diag}\left(\bm{\pi}_{2,i}^{*}\right)\mathbf{F}_{k,2}\operatorname{diag}\left(\bm{\pi}_{2,i}\right)$, and $\mathbf{R} \triangleq \sum_{k\in\mathcal{K}}\widebar{\bm{\Pi}}_{1,k} + \delta_{1}^{2}\mathbf{I}_{N_{1}}$. Then, by fixing $\left(\mathbf{w},\bm{\phi}_{2},\bm{\mu},\bm{\nu}\right)$ and dropping the respective constant terms from the OF, we obtain the following sub-problem for optimizing $\bm{\phi}_{1}$:
\begin{equation}\label{eq:36}
    \text{(P4-A): } \underset{\bm{\phi}_{1}}{\max} \operatorname{Re}\left\{\bm{\phi}_{1}^{\dagger}\bm{\upsilon}_{1}\right\}-\bm{\phi}_{1}^{\dagger}\mathbf{Q}_{1}\bm{\phi}_{1} \ \text{s.t. }\text{C2: } \bm{\phi}_{1}^{\dagger}\mathbf{R}\bm{\phi}_{1} \leq \widetilde{P}_{1}^{\max}.
\end{equation}
We use the Lagrange multipliers method to tackle this standard convex QCQP and obtain a closed-form expression of $\bm{\phi}_{1}^{\star}$:
\begin{equation}\label{eq:37}
    \bm{\phi}_{1}^{\star} = \frac{1}{2}\left(\mathbf{Q}_{1}+\varpi\mathbf{R}\right)^{-1}\bm{\upsilon}_{1},
\end{equation}
where the Lagrange multiplier associated with C2, $\varpi$, is optimized via binary search. 

Next, with fixed $\left(\mathbf{w},\bm{\phi}_{1},\bm{\mu},\bm{\nu}\right)$, we form the minimization sub-problem:
\begin{equation}\label{eq:38}
    \text{(P4-B): } \underset{\bm{\phi}_{2}}{\min} \ \bm{\phi}_{2}^{\dagger}\mathbf{Q}_{2}\bm{\phi}_{2} - \operatorname{Re}\left\{\bm{\phi}_{2}^{\dagger}\bm{\upsilon}_{2}\right\} \ \text{s.t. C3}.
\end{equation}
We apply the MM method to handle the UMCs C3~\cite{MM}. Specifically, for any given solution $\bm{\phi}_{2}^{(t)}$ at the $t$-th iteration of the MM algorithm and any feasible $\bm{\phi}_{2}$, we have $\bm{\phi}_{2}^{\dagger}\mathbf{Q}_{2}\bm{\phi}_{2} \leq \bm{\phi}_{2}^{\dagger}\mathbf{X}\bm{\phi}_{2}-2\operatorname{Re}\left\{\bm{\phi}_{2}^{\dagger}\left(\mathbf{X}-\mathbf{Q}_{2}\right)\bm{\phi}_{2}^{(t)}\right\}+\left(\bm{\phi}_{2}^{(t)}\right)^{\dagger}\left(\mathbf{X}-\mathbf{Q}_{2}\right)\bm{\phi}_{2}^{(t)}\triangleq y\left(\bm{\phi}_{2}|\bm{\phi}_{2}^{(t)}\right)$, where $\mathbf{X}\in\mathbb{R}_{+}^{N_{2}\times N_{2}}$ is defined as $\mathbf{X}\triangleq\lambda_{Q}\mathbf{I}_{N_{2}}\succeq\mathbf{Q}_{2}$ and $\lambda_{Q}=\lambda_{\max}\left(\mathbf{Q}_2\right)$---alternatively, we can set $\lambda_Q=\operatorname{Tr}\left(\mathbf{Q}_2\right)$ to reduce the computational complexity from $\mathcal{O}\left(N_2^3\right)$ to $\mathcal{O}\left(N_2\right)$. Thus, we replace the OF in (P4-B) by a surrogate OF, $z\left(\bm{\phi}_{2}|\bm{\phi}_{2}^{(t)}\right)\triangleq y\left(\bm{\phi}_{2}|\bm{\phi}_{2}^{(t)}\right)-\operatorname{Re}\left\{\bm{\phi}_{2}^{\dagger}\bm{\upsilon}_{2}\right\}$. By removing the constant terms in this OF (e.g., $\bm{\phi}_{2}^{\dagger}\mathbf{X}\bm{\phi}_2=N_{2}\lambda_{Q}$), we recast (P4-B) as:  
\begin{equation}\label{eq:40}
    \text{(P4-C): } \underset{\bm{\phi}_{2}}{\max} \ \operatorname{Re}\left\{\bm{\phi}_{2}^{\dagger}\mathbf{q}^{(t)}\right\} \ \text{s.t. C3},
\end{equation}
where $\mathbf{q}^{(t)}\in\mathbb{C}^{N_{2}}$ is defined as $\mathbf{q}^{(t)}\triangleq\left(\mathbf{X}-\mathbf{Q}_{2}\right)\bm{\phi}_{2}^{(t)}+\bm{\upsilon_{2}}$. The optimal solution of (P4-C) is given by
\begin{equation}\label{eq:41}
    \bm{\phi}_{2}^{\star}=\bm{\phi}_{2}^{(t+1)} = e^{j\operatorname{arg}\left(\mathbf{q}^{(t)}\right)}.
\end{equation}

\subsection{Hybrid Active RIS}\label{subsec:4.2}
In the hybrid active RIS case, the optimization problem of interest is formulated as follows: 
\begin{subequations}\label{eq:41add}
\begin{alignat}{2}
&&&\text{(P5): }\underset{\mathbf{w},\left\{\bm{\phi}_{s}\right\}}{\max} \ \eta=\frac{R}{P} = \frac{\sum\limits_{k\in\mathcal{K}}\log_{2}\left(1+\gamma_{k}\right)}{\widetilde{P}+\sum_{s\in\mathcal{S}}P_{s}} \label{eq:41adda} \\
&&&\text{s.t.} \ \ \ \ \text{C5: }P_{\text{BS}} = \xi^{-1}\sum_{k\in\mathcal{K}}\left\|\mathbf{w}_{k}\right\|^{2} + W_{\text{BS}} \leq P_{\text{BS}}^{\max}, \label{eq:41addb} \\
&&& \ \ \ \ \ \ \ \text{C6: }P_{s} = \zeta_{s}^{-1}\left(\sum_{k\in\mathcal{K}}\left\|\bm{\Phi}_{s}\mathbf{G}_{s}\mathbf{w}_{k}\right\|^{2} + \delta_{s}^{2}\left\|\bm{\Phi}_{s}\right\|_{F}^{2}\right) \nonumber \\
&&& \ \ \ \ \ \ \ \ \ \ \ \ \ \ +W_{r,s} \leq P_{s}^{\max}, \forall s\in\mathcal{S}, \label{eq:41addc}
\end{alignat}
\end{subequations}
where $\gamma_{k}$ is given by Eq.~(\ref{eq:4}), $W_{r,1}=N_{1}\left(P_{\text{PS}}+P_{\text{DC}}\right)$ (FC) or $W_{r,1}=N_{1}P_{\text{PS}}+L_1 P_{\text{DC}}$ (SC), and $W_{r,2}=N_{2}P_{\text{PS}}+L_2 P_{\text{DC}}$.

\subsubsection{Fractional Programming}\label{subsubsec:4.2.1}
By using Dinkelbach's algorithm along with the Lagrangian dual and quadratic transforms~\cite{FP1}, we equivalently recast (P5) as:
\begin{equation}\label{eq:42}
    \text{(P6): }\underset{\mathbf{w},\bm{\phi},\bm{\mu}^{\prime},\bm{\nu}^{\prime}}{\max} \ g^{\prime}\left(\mathbf{w},\bm{\phi},\bm{\mu}^{\prime},\bm{\nu}^{\prime}\right) \ \text{s.t. C5, C6},
\end{equation}
where $\bm{\mu}^{\prime}\in\mathbb{C}^{K}$ and $\bm{\nu}^{\prime}\in\mathbb{C}^{K}$ are auxiliary variables, $g^{\prime}\left(\mathbf{w}, \bm{\phi}, \bm{\mu}^{\prime}, \bm{\nu}^{\prime}\right) = -\eta P +\sum_{k\in\mathcal{K}}\ln \left(\widebar{\mu}_k^{\prime}\right)-\mu_k^{\prime} +2 \sqrt{\widebar{\mu}_k^{\prime}}\rho_k^{\prime} -\sum_{k\in\mathcal{K}}\left|\nu_k^{\prime}\right|^2\left(\sum_{i\in\mathcal{K}}\left|\mathbf{h}_{k}^{\dagger}\mathbf{w}_i\right|^2+\sum_{s\in\mathcal{S}}\left\|\mathbf{f}_{k,s}^{\dagger} \bm{\Phi}_{s}\right\|^2 \delta_s^2+\sigma_{k}^2\right)$, $\widebar{\mu}_k^{\prime}\triangleq 1+\mu_k^{\prime}$, and $\rho_{k}^{\prime}\triangleq\operatorname{Re}\left\{\left(\nu_{k}^{\prime}\right)^{*}\mathbf{h}_{k}^{\dagger}\mathbf{w}_{k}\right\}$. Next, we develop a BCA algorithm to alternately optimize the auxiliary and decision variables, as described below.

\subsubsection{Blocks Update}\label{subsubsec:4.2.2}
\paragraph{Update Auxiliary Variables}
Setting $\partial g^{\prime}/\partial \mu_{k}^{\prime}=0$ and $\partial g^{\prime}/\partial \nu_{k}^{\prime}=0$ with fixed $\left(\mathbf{w},\bm{\phi},\bm{\nu}^{\prime}\right)$ and $\left(\mathbf{w},\bm{\phi},\bm{\mu}^{\prime}\right)$, respectively, we obtain
\begin{subequations}\label{eq:44}
\begin{alignat}{2}
\mu_k^{\prime,\star} & =\frac{\rho_{k}^{\prime}}{2}\left(\rho_k^{\prime}+\sqrt{\left(\rho_k^{\prime}\right)^{2}+4}\right), \label{eq:44b} \\
\nu_k^{\prime,\star} & =\frac{\sqrt{\widebar{\mu}_k^{\prime}} \mathbf{h}_{k}^{\dagger} \mathbf{w}_k}{\sum\limits_{i\in\mathcal{K}}\left|\mathbf{h}_{k}^{\dagger} \mathbf{w}_i\right|^2+\sum\limits_{s\in\mathcal{S}}\left\|\mathbf{f}_{k,s}^{\dagger} \bm{\Phi}_{s}\right\|^2 \delta_s^2+\sigma_{k}^2}. \label{eq:44a}
\end{alignat}
\end{subequations}

\paragraph{Update Transmit Precoding}
From Eq.~(\ref{eq:41addc}),
\begin{equation}\label{eq:45}
    \widetilde{P}_{2}^{\max} = \zeta_{2}\left(P_{2}^{\max}-W_{r,2}\right). 
\end{equation}
We also respectively define $\mathbf{S}^{\prime}\in\mathbb{C}^{KM\times KM}$ and $\mathbf{T}_{2}\in\mathbb{C}^{KM\times KM}$ as
\begin{equation}\label{eq:46}
 \mathbf{S}^{\prime} \triangleq \mathbf{I}_{K}\otimes\left(\eta\xi^{-1}\mathbf{I}_{M}+\sum_{s\in\mathcal{S}}\eta\zeta_{s}^{-1}\mathbf{E}_{s}^{\dagger}\mathbf{E}_{s}+\sum_{k\in\mathcal{K}}\left|\nu_k^{\prime}\right|^2\mathbf{H}_{k}\right),
\end{equation}
and $\mathbf{T}_2 \triangleq \mathbf{I}_{K}\otimes \left(\mathbf{E}_{2}^{\dagger}\mathbf{E}_{2}\right)$, where $\mathbf{E}_s\triangleq\bm{\Phi}_s\mathbf{G}_s$. Hence, with fixed $\left(\bm{\phi},\bm{\mu}^{\prime},\bm{\nu}^{\prime}\right)$, the transmit precoding optimization sub-problem can be formulated as follows:
\begin{subequations}\label{eq:47}
    \begin{alignat}{2}
        &&&\text{(P7): } \underset{\mathbf{w}}{\max} \operatorname{Re}\left\{\mathbf{u}^{\dagger}\mathbf{w}\right\}-\mathbf{w}^{\dagger}\mathbf{S}^{\prime}\mathbf{w} \label{eq:47a} \\
        &&&\text{s.t.} \ \ \ \ \text{C5: }\mathbf{w}^{\dagger}\mathbf{w} \leq \widetilde{P}_{\text{BS}}^{\max}, \label{eq:47b} \\
        &&& \ \ \ \ \ \ \ \text{C6: } \mathbf{w}^{\dagger}\mathbf{T}_{s}\mathbf{w} \leq \widetilde{P}_{s}^{\max} - \delta_{s}^{2}\left\|\bm{\Phi}_{s}\right\|_{F}^{2}, \ \forall s\in\mathcal{S}. \label{eq:47c}
    \end{alignat}
\end{subequations}
By using the Lagrange multipliers method to tackle this standard convex QCQP, we obtain:
\begin{equation}\label{eq:48}
    \mathbf{w}^{\star} = \frac{1}{2}(\mathbf{S}^{\prime}+\lambda^{\prime}\mathbf{I}_{KM}+\sum_{s\in\mathcal{S}}\psi_{s}\mathbf{T}_{s})^{-1}\mathbf{u},
\end{equation}
where $\lambda^{\prime}$ and $\psi_{s}$ represent the Lagrange multipliers, which are associated with constraints C5 and C6, respectively, and are optimized via grid search.

\paragraph{Update RIS Beamforming}
Let us define $\mathbf{Q}_{s}^{\prime}\in\mathbb{C}^{N_{s}\times N_{s}}$ and $\mathbf{R}_{s}\in\mathbb{C}^{N_{s}\times N_{s}}$ as
\begin{align}\label{eq:49}
\mathbf{Q}_{s}^{\prime} \triangleq & \sum_{k\in\mathcal{K}}\left|\nu_{k}^{\prime}\right|^{2}\left(\delta_{s}^{2}\widebar{\mathbf{F}}_{k,s}+\sum_{i\in\mathcal{K}}\operatorname{diag}\left(\bm{\pi}_{s,i}^{*}\right)\mathbf{F}_{k,s}\operatorname{diag}\left(\bm{\pi}_{s,i}\right)\right)\nonumber \\
        &+\sum_{k\in\mathcal{K}}\eta\zeta_{s}\left(\widebar{\bm{\Pi}}_{s,k}+\delta_{s}^{2}\mathbf{I}_{N_{s}}\right), 
\end{align}
and $\mathbf{R}_s \triangleq \sum_{k\in\mathcal{K}}\widebar{\bm{\Pi}}_{s,k} + \delta_{s}^{2}\mathbf{I}_{N_{s}}$, respectively, where $\widebar{\mathbf{F}}_{k,s}\triangleq\operatorname{diag}\left(\mathbf{f}_{k, s} \odot \mathbf{f}_{k, s}^*\right)\in\mathbb{C}^{N_s\times N_s}$, and $\widebar{\bm{\Pi}}_{s,k}\triangleq\operatorname{diag}\left(\boldsymbol{\pi}_{s, k} \odot \boldsymbol{\pi}_{s, k}^*\right)\in\mathbb{C}^{N_s\times N_s}$. Hence, with fixed $\left(\mathbf{w},\bm{\mu}^{\prime},\bm{\nu}^{\prime}\right)$, we can formulate the RIS beamforming optimization sub-problem as follows:
\begin{subequations}\label{eq:50}
\begin{alignat}{2}
&&&\text{(P8): } \underset{\bm{\phi}}{\max} \sum_{s\in\mathcal{S}}\left(\operatorname{Re}\left\{\bm{\phi}_{s}^{\dagger}\bm{\upsilon}_{s}\right\}-\bm{\phi}_{s}^{\dagger}\mathbf{Q}_{s}^{\prime}\bm{\phi}_{s}\right) \label{eq:50a} \\
        &&&\text{s.t.} \ \ \ \ \text{C6: } \bm{\phi}_{s}^{\dagger}\mathbf{R}_{s}\bm{\phi}_{s} \leq \widetilde{P}_{s}^{\max}.\label{eq:50b} 
\end{alignat}
\end{subequations}
Fixing $\left(\mathbf{w},\bm{\phi}_{s^{\prime}},\bm{\mu},\bm{\nu}\right)$ and solving for $\bm{\phi}_{s}$, $s,s^{\prime}\in\mathcal{S}$, $s^{\prime}\neq s$, we formulate respective standard convex QCQPs. By using the Lagrange multipliers method, we obtain:
\begin{equation}\label{eq:51}
    \bm{\phi}_{s}^{\star} = \frac{1}{2}\left(\mathbf{Q}_{s}^{\prime}+\varpi_{s}\mathbf{R}_{s}\right)^{-1}\bm{\upsilon}_{s},
\end{equation}
where $\varpi_{s}$ represents the Lagrange multiplier associated with the constraints C6 with fixed $s=1$ or $s=2$ in each case, respectively, and is optimized via binary search. 
 \begin{algorithm}[!t]
 \centering \scriptsize
   \begin{algorithmic}[1]
   \State Randomly initialize $\mathbf{w}$ and $\bm{\phi}_{s}$, $s\in\mathcal{S}$; set $j=0$; set $T_{\max}$.
   \Repeat
   \State Update $\bm{\mu}$ and $\bm{\nu}$ via Eq.~(\ref{eq:29})/$\bm{\mu}^{\prime}$ and $\bm{\nu}^{\prime}$ via Eq.~(\ref{eq:44}).
   \State Update $\mathbf{w}$ via Eq.~(\ref{eq:34})/Eq.~(\ref{eq:48}).
   \State Update $\bm{\phi}_{1}$ via Eq.~(\ref{eq:37})/Eq.~(\ref{eq:51}).
   \State Update $\bm{\phi}_{2}$ via Eq.~(\ref{eq:41})/Eq.~(\ref{eq:51}).
   \State $j \leftarrow j+1$
   \Until Convergence or $T_{\max}$ iterations are reached.
   \State \textbf{Output:} $\left\{\mathbf{w}^{\star},\bm{\phi}_{s}^{\star}\right\}$.
 \end{algorithmic}
 \caption{BCA Algorithm for Solving (P2)/(P6).}
 \label{algo:1}
 \end{algorithm}
\begin{figure*}[!t]
\centering
\subfloat[]{
	\label{subfig:new1}
	\includegraphics[scale = 0.3]{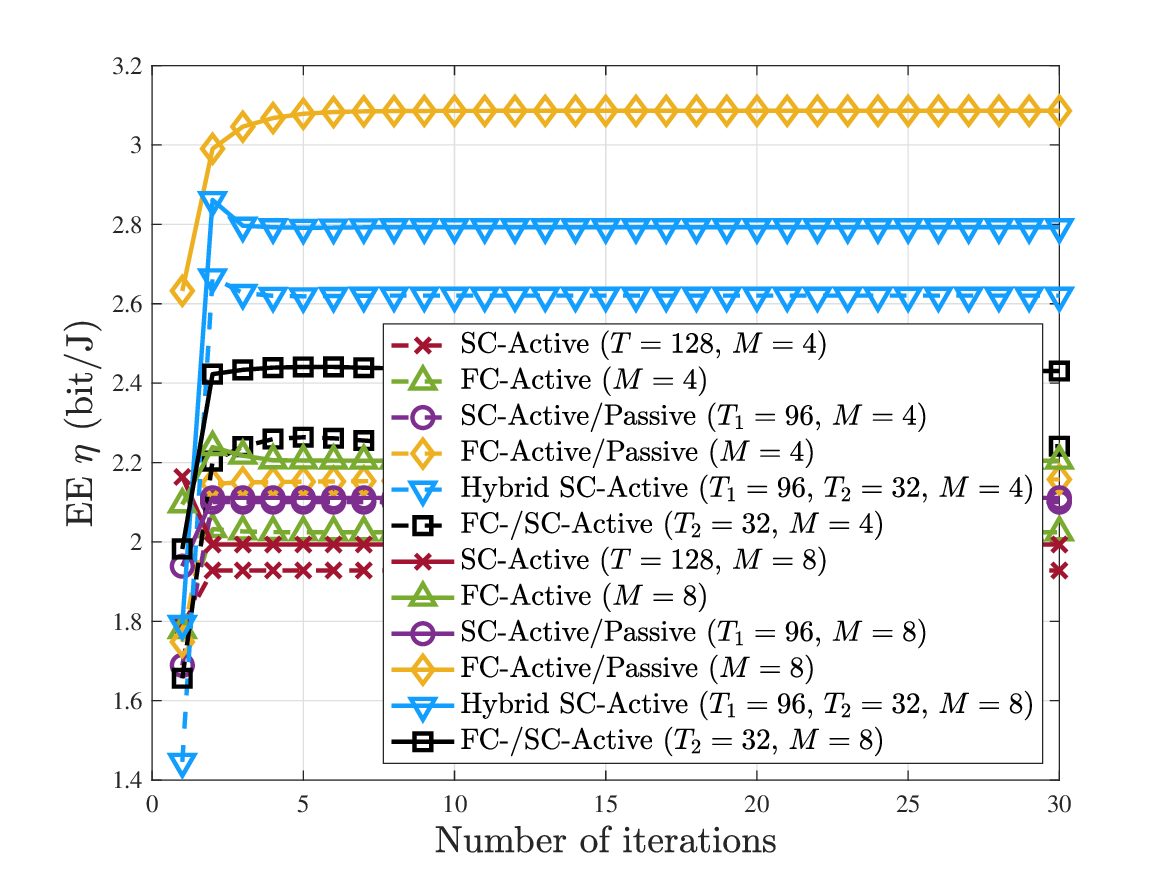} } 
\subfloat[]{
	\label{subfig:new2}
	\includegraphics[scale = 0.3]{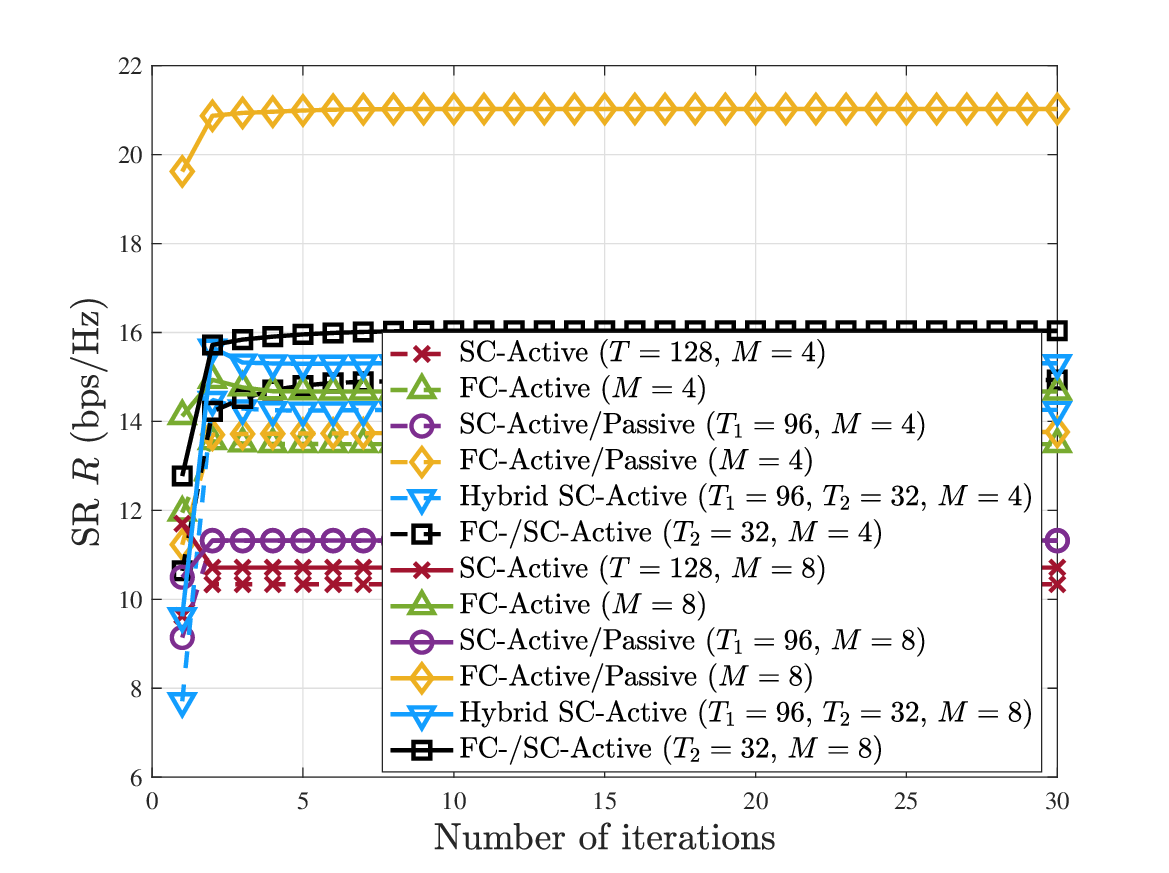} } 
\caption{EE $\eta$ (left) and SR $R$ (right) vs. the number of iterations for $M=4$ or $M=8$. ($N=128$, $a=0.75$, $L=1$, $D=300$ m.).}
\label{fig:new}
\end{figure*}
\begin{figure*}[!t]
\centering
\subfloat[]{
	\label{subfig:5a}
	\includegraphics[scale = 0.3]{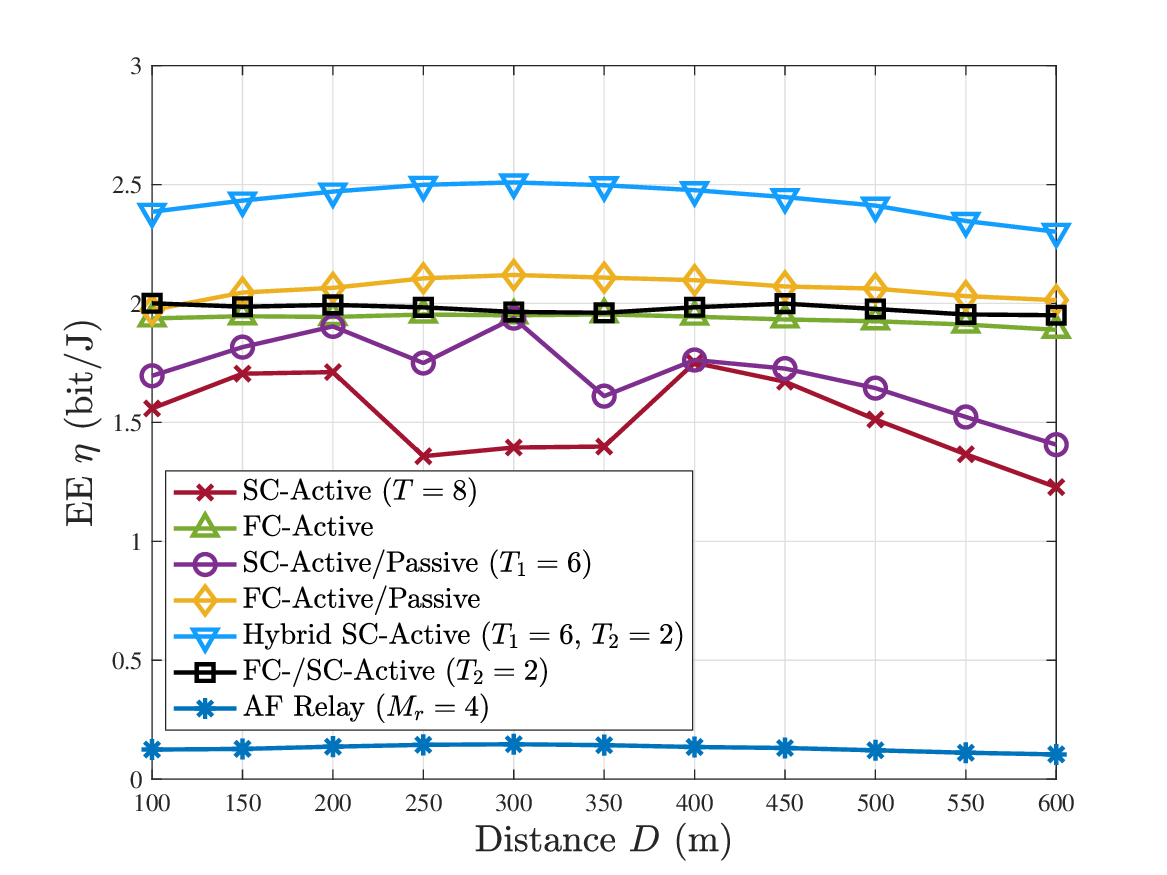} } 
	\subfloat[]{
	\label{subfig:5b}
	\includegraphics[scale = 0.3]{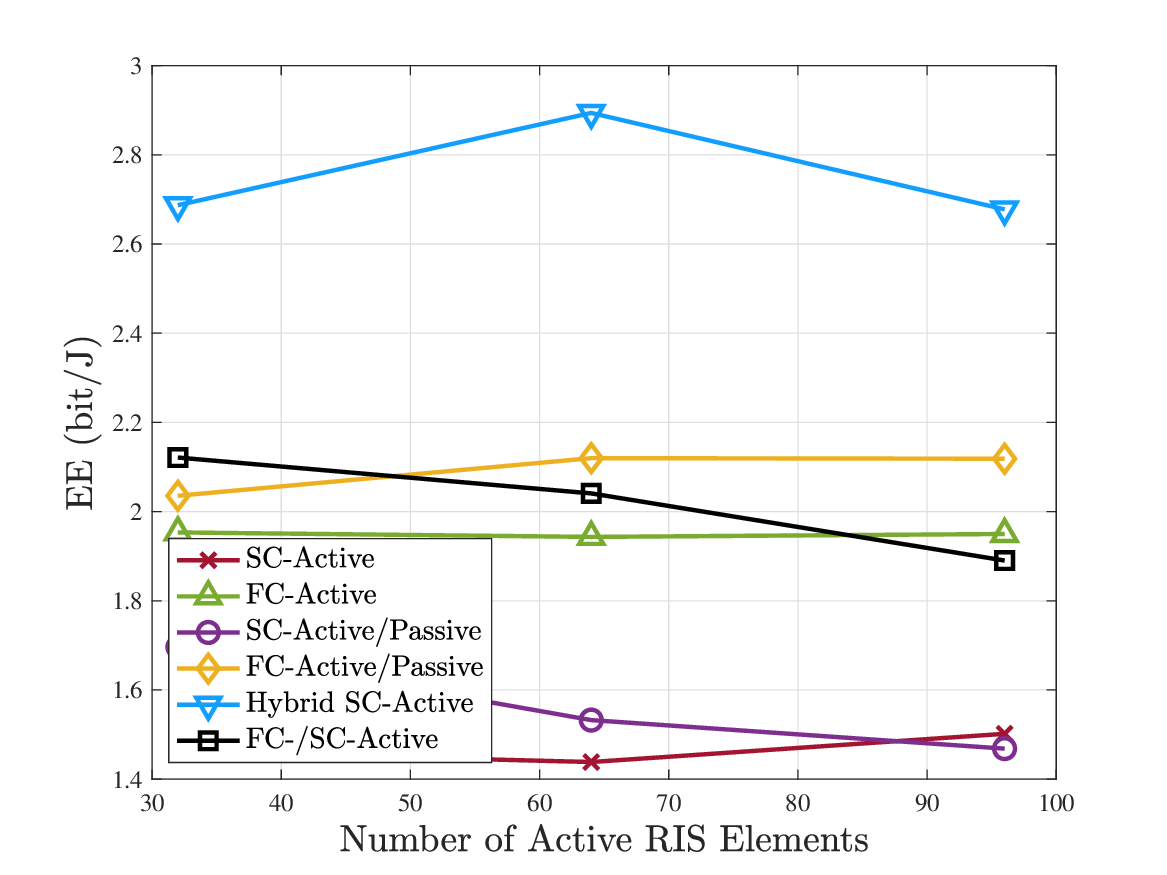} } 
	\subfloat[]{
	\label{subfig:5c}
	\includegraphics[scale = 0.3]{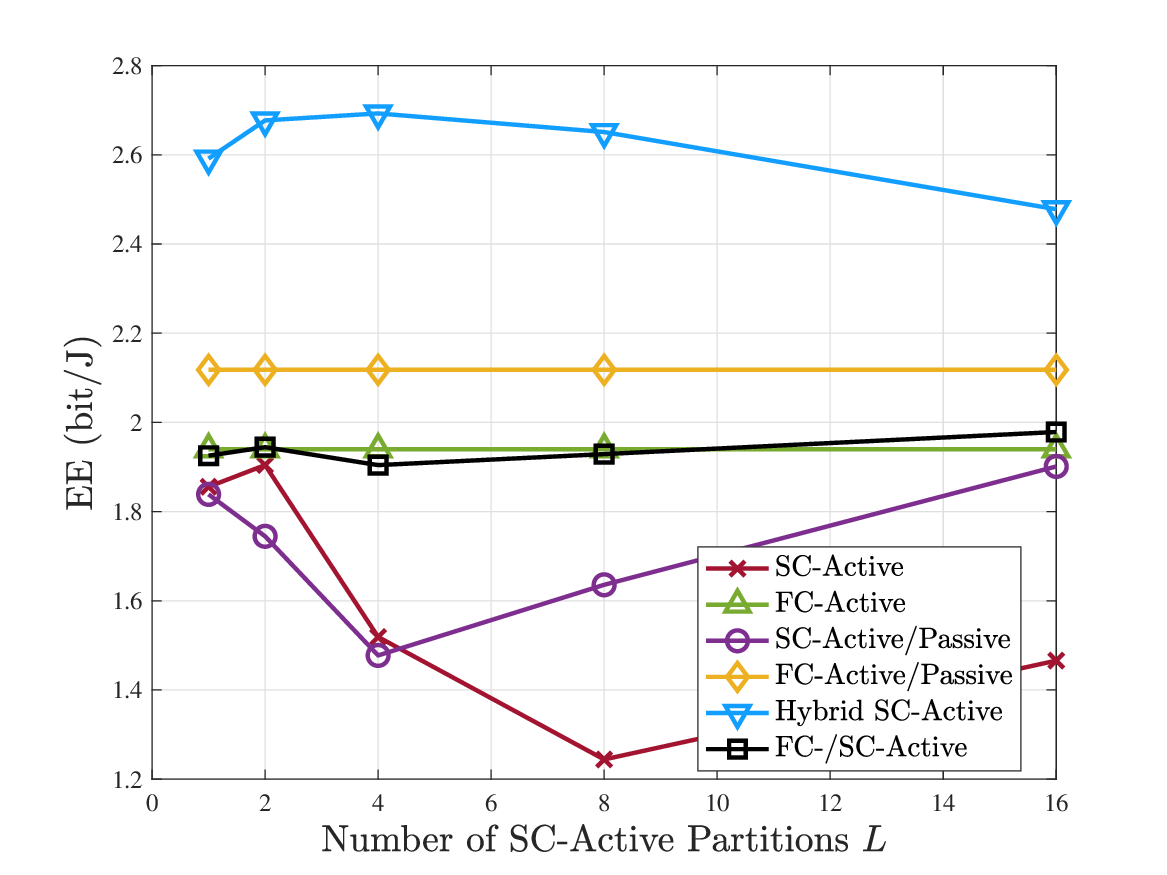} } 
	
\subfloat[]{
	\label{subfig:5d}
	\includegraphics[scale = 0.3]{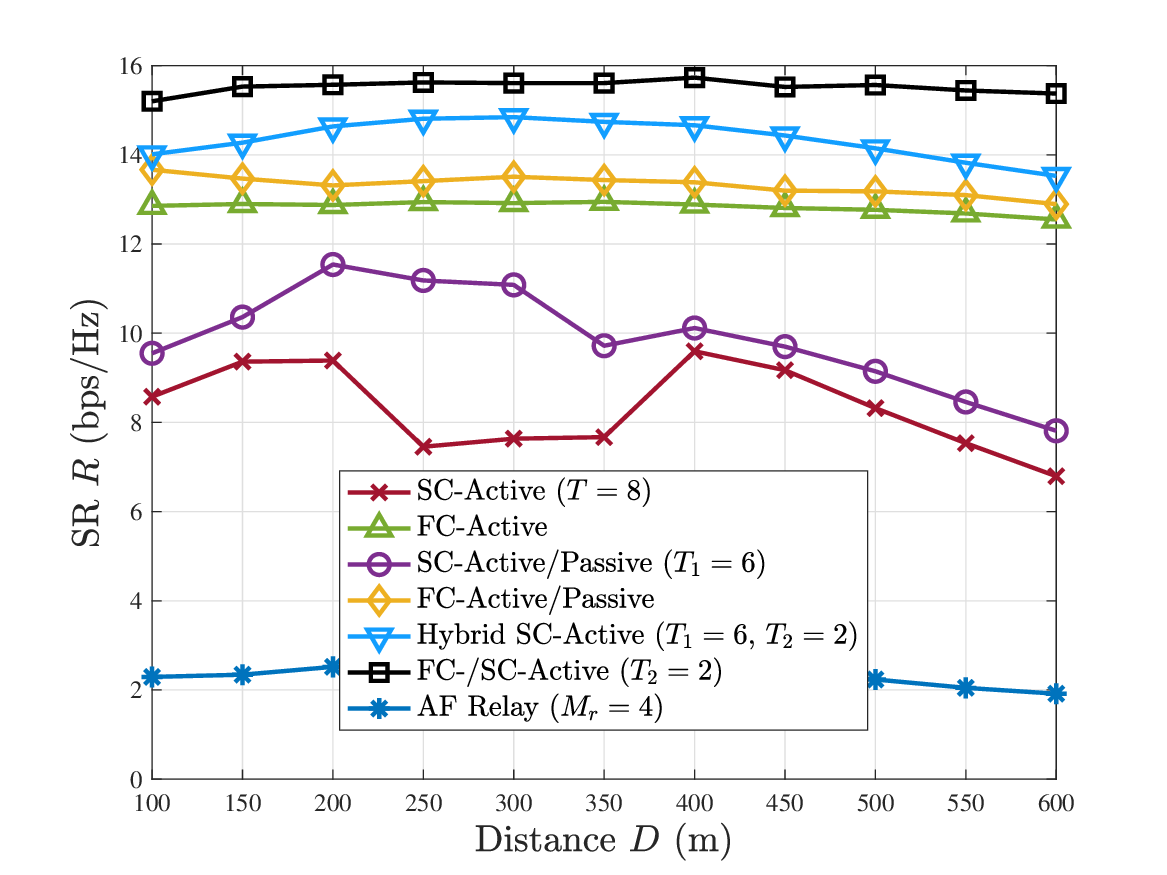} } 
	\subfloat[]{
	\label{subfig:5e}
	\includegraphics[scale = 0.3]{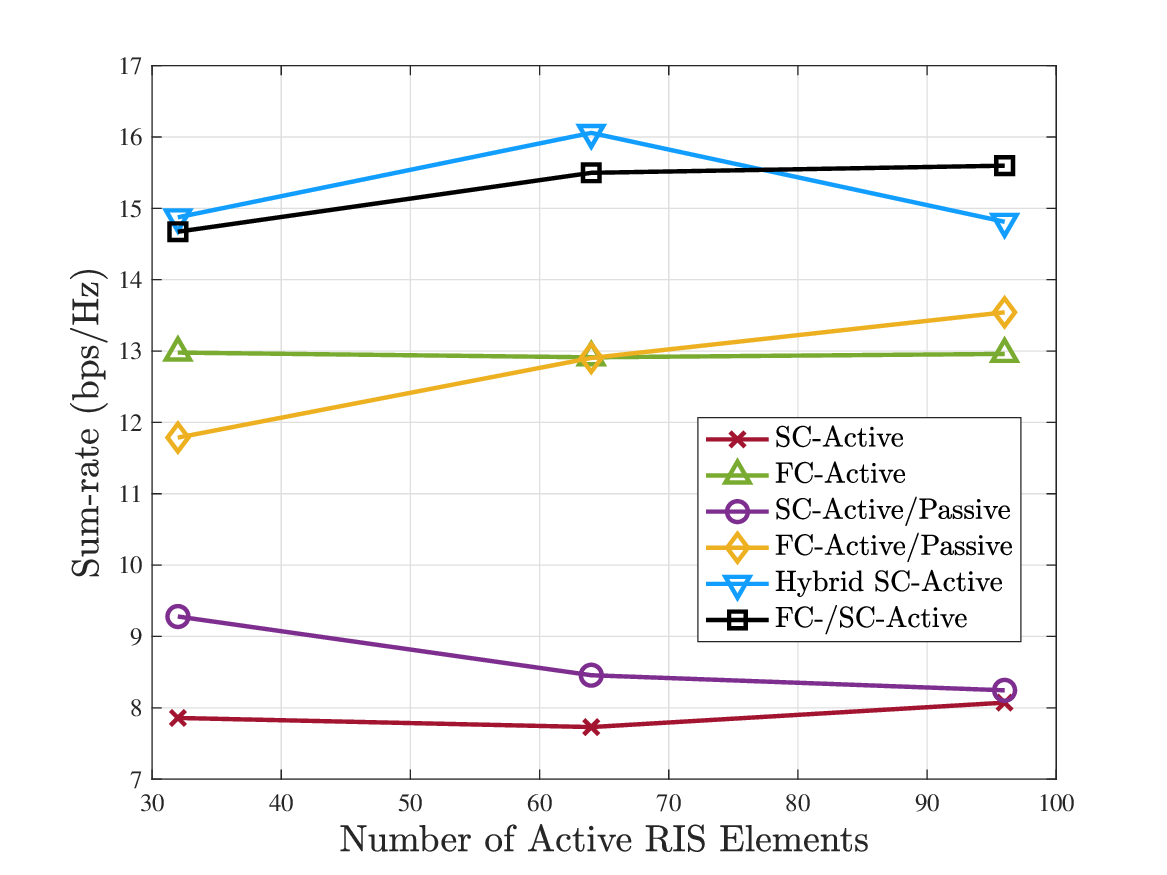} } 
	\subfloat[]{
	\label{subfig:5f}
	\includegraphics[scale = 0.3]{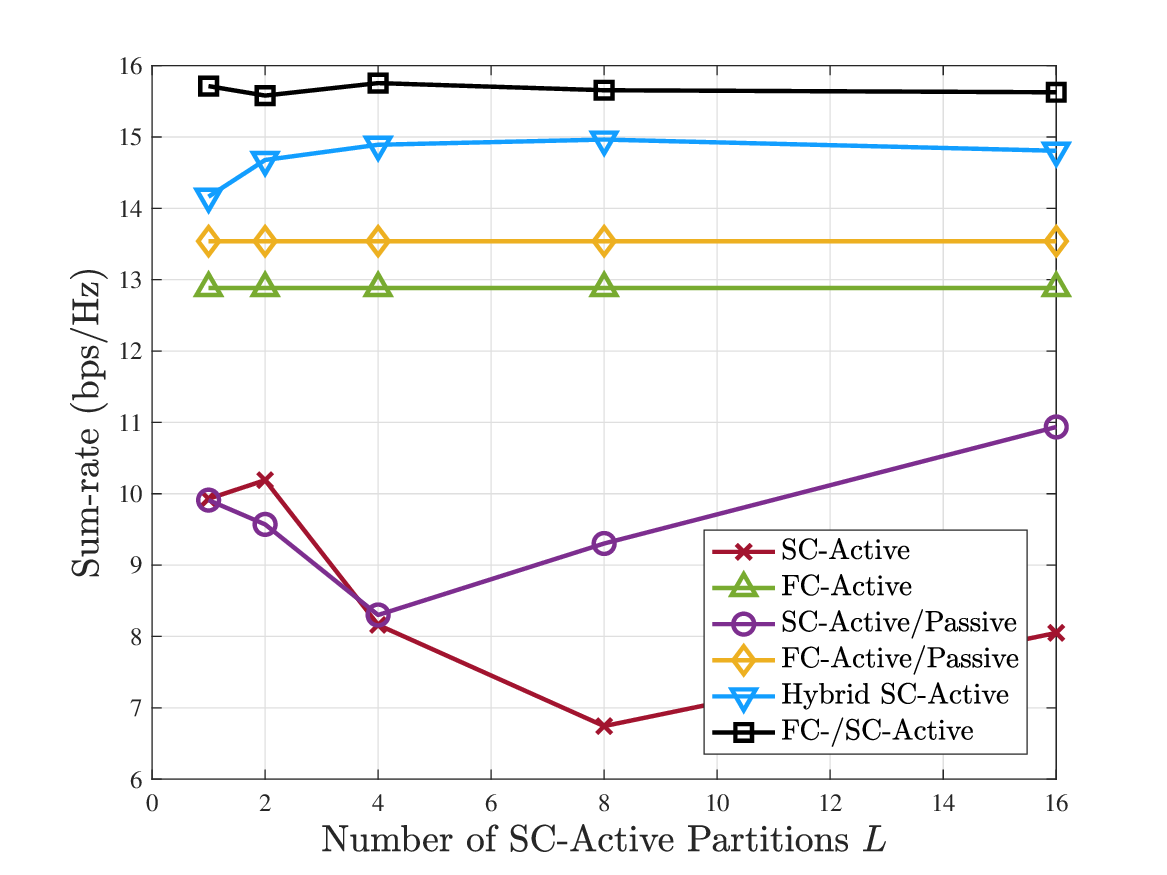} } 
\caption{EE $\eta$ (top) and SR $R$ (right) vs. $D=100:50:600$ m for $N=128$, $a=0.75$, $L=4$ (left), $a=0.125:0.125:0.875$ for $N=128$, $L=4$, $D=300$ m (center), $L\in\{1,2,4,8,16\}$ for $N=128$, $a=0.75$, $D = 300$ m (right).}
\label{fig:5newa}
\end{figure*}

\subsection{Complexity Analysis}\label{subsec:4.3}
The BCA algorithm for solving problem (P2) or (P6) is presented in Algorithm~\ref{algo:1}. The combined complexity of (P3) and (P4-A) is given by $\mathcal{C}_{1}=\mathcal{O}\left(\log _2(1 / \varepsilon) \left(M^{4.5} K^{4.5}+N_{1}^{4.5}\right)\right)$, while the complexity of (P4-B) is $\mathcal{C}_{2}=\mathcal{O}\left(N_{2}^{3}+I_{MM}N_{2}^{2}\right)$, where $I_{MM}$ denotes the number of iterations required for converge of the MM algorithm~\cite{MajMin}. Likewise, the combined complexity of (P7) and (P8) is given by $\mathcal{C}_{3}=\mathcal{O}\left(\log _2(1 / \varepsilon) \left(M^{4.5} K^{4.5}+N_{1}^{4.5}+N_{2}^{4.5}\right)\right)$. The complexity of updating the auxiliary variables equals $\mathcal{O}\left(KM\right)$ for $\bm{\mu}$ and $\bm{\mu}^{\prime}$, $\mathcal{O}\left(K^{2}M+KN_{1}\right)$ for $\bm{\nu}$, and $\mathcal{O}\left(K^{2}M+KN\right)$ for $\bm{\nu}^{\prime}$. Therefore, the overall complexity of the developed BCA algorithm for the active/passive and active/active RIS cases is respectively given by $\mathcal{O}\left(I_{0}\max\left\{\mathcal{C}_{1},\mathcal{C}_{2}\right\}\right)$ and $\mathcal{O}\left(\log _2(1 / \varepsilon) I_{0}^{\prime}\left(M^{4.5} K^{4.5}+N_{1}^{4.5}+N_{2}^{4.5}\right)\right)$, where $I_{0}$, $I_{0}^{\prime}$ denote the number of iterations required for the algorithm to terminate, i.e. converge or reach $T_{\max}$ iterations.

\section{Numerical Evaluations}\label{sec:5}
In this section, we comparatively evaluate the EE and SR of the joint precoding/RIS beamforming designs proposed in Sec.~\ref{sec:4} against benchmarks via numerical simulations. While our optimization objective focuses on maximizing EE, we have also included SR results in our simulations for several important reasons. First, the combined presentation of EE and SR trends enables us to understand the trade-off between energy consumption and performance. Second, in practical deployments, network operators often need to balance both EE and SR requirements, making it beneficial to present both metrics side-by-side. Third, the inclusion of SR results allows for a more direct comparison with existing studies that may focus solely on that performance metric. Lastly, by presenting both EE and SR, we offer a more comprehensive view of how different hybrid RIS architectures perform across multiple key performance indicators. Thus, this dual presentation of results enhances the overall utility and impact of our work.

In the  simulations, we consider a setting where the coordinates of the BS and the RIS (in meters) are $(0,-60)$ and $(300,10)$, while the  users are randomly distributed on a circular cluster with radius $r=5$ m and center at $(D,0)$. The simulation parameters are $N=128$, $M=4$, $K=2$, $W_{\text{BS}} = 6$ dBW, $W_{\text{PS}} = W_{\text{DC}}=10$ dBm, $P_{\text{BS}}^{\max}=P_{s }^{\max}=9$ dBW, $\xi=\zeta_{s}=0.909$, and $\sigma_{k}^{2}=\delta_{s}^{2}=-80$ dBm, $\forall s\in\mathcal{S}$, $\forall k\in\mathcal{K}$~\cite{Subconnected}, unless mentioned otherwise. Furthermore, Rician fading channels are considered, with Rician factor $\kappa=0$ dB. The path loss is given by $\text{PL}(d)=37.3+22.0\log_{10}d$ for the BS--RIS and RIS--user links or $\text{PL}(d)=41.2+28.7\log_{10}d$ for the BS--user links, where $d$ is the propagation distance~\cite{IRSAct8}.

\textbf{Convergence of the Optimization Algorithm and Impact of the Number of Antennas:} Algorithm 1 converges within 5 iterations across various system configurations, as shown in Fig.~\ref{fig:new}. Increasing the number of transmit antennas improves both the EE and SR for all RIS architectures, due to the enhanced beamforming and interference management capabilities. FC-active/passive RIS presents the best EE and SR at $M=8$, thus efficiently exploiting the increased spatial degrees-of-freedom of the BS, but performs poorly at $M=4$ where the BS gains are lower, as the passive RS reduces the SR, and consequently the EE, in this regime.

\textbf{Performance Trends across Architectures:} FC-active/SC-active RIS achieves the highest SR, especially at moderate to large RIS sizes, via the flexible combination of fine-grained signal control and power-efficient beamforming. Hybrid SC-active RIS exhibits the best EE across different scenarios through SR/TPC balancing via non-uniform partitioning and amplifier sharing. FC-active RIS demonstrates lower EE due to the high power consumption attributed to the large number of amplifiers, as well as diminishing SR in this scenario caused by the increased amplification noise. SC-active/passive and SC-active RIS show the worst SR and EE performance, due to the limited number of power amplifiers.

\textbf{Influence of User Location and Comparison with Relays:} As shown in the left column of Fig.~\ref{fig:5newa}, the EE and SR of SC-active, hybrid SC-active, and SC-active/passive RIS architectures peak when the users are located near the RIS. All RIS variants significantly outperform a relay with 4 RF-fed antennas that employs the amplify-and-forward (AF) scheme. This stems from RIS's precise beamforming control and channel shaping ability, efficient power usage, and controlled noise amplification (contrary to AF relay) caused by the use of 128 reflecting elements and the sharing of reflect-type power amplifiers among groups of 32, 24, or 8 RIS elements.

\textbf{Impact of RIS Elements Allocation:} The central column of Fig.~\ref{fig:5newa} shows that the performance of hybrid SC-active RIS peaks with equal elements allocation ($a=0.5$), since then the effects of increased active RSs count and respectively reduced RS's power budget are balanced. FC-active/passive RIS's performance improves with an increase of $a$, as the SR gains attributed to the higher number of active elements outweigh the TPC increase. SC-active/passive RIS's performance declines with $a$, demonstrating passive RS's effectiveness with limited number of amplifiers as in Sec.~\ref{sec:3}. FC-active/SC-active RIS shows declining EE but improving SR with $a$, as expected.

\textbf{Impact of the Number of RIS Partitions:} In the right column of Fig.~\ref{fig:5newa}, we notice that hybrid SC-active RIS's EE peaks at $L=4$ before declining due to the increased TPC. SC-active/passive RIS's EE drops until $L=4$ and then grows, while SC-active RIS's EE peaks at $L=2$, drops until $L=8$, and then increases as the SR gains offset the TPC penalties.

\begin{figure*}[!t]
\centering
\subfloat[]{
	\label{subfig:6a}
	\includegraphics[scale = 0.3]{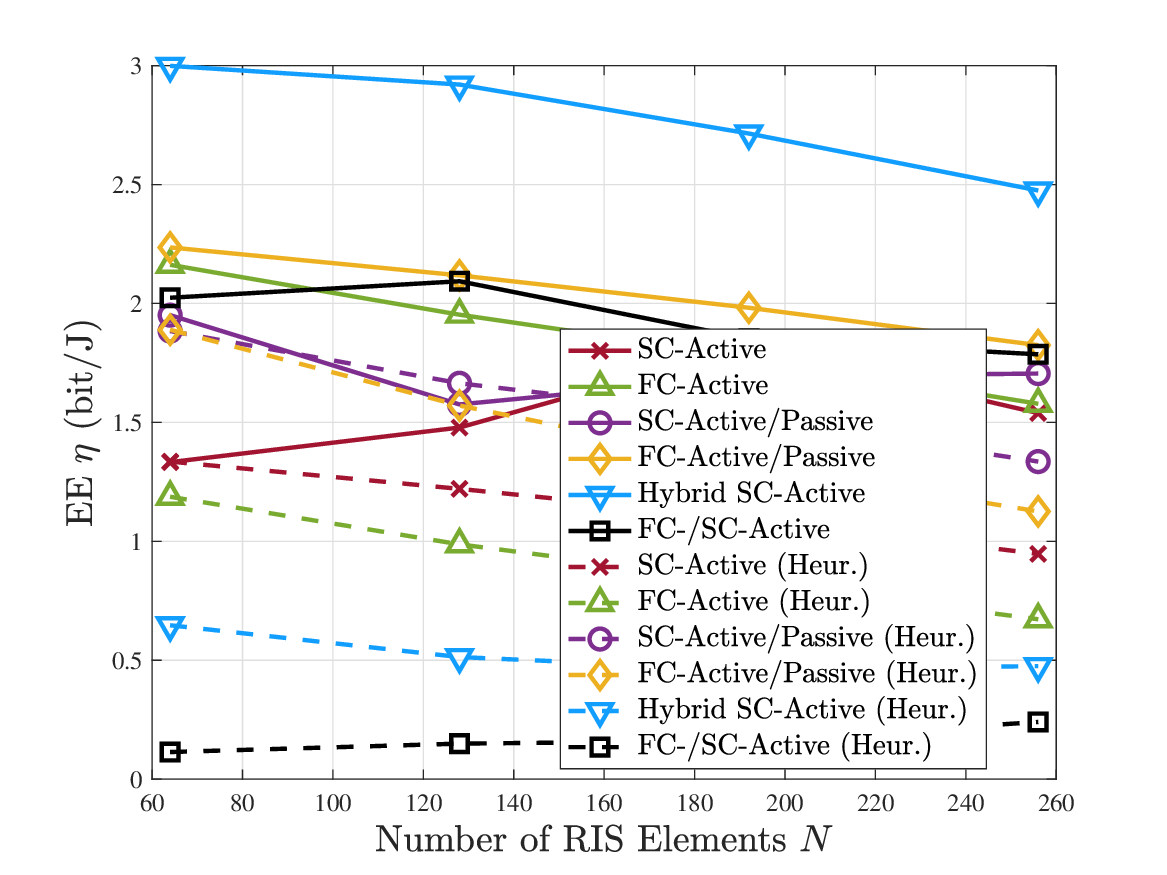} } 
	\subfloat[]{
	\label{subfig:6b}
	\includegraphics[scale = 0.3]{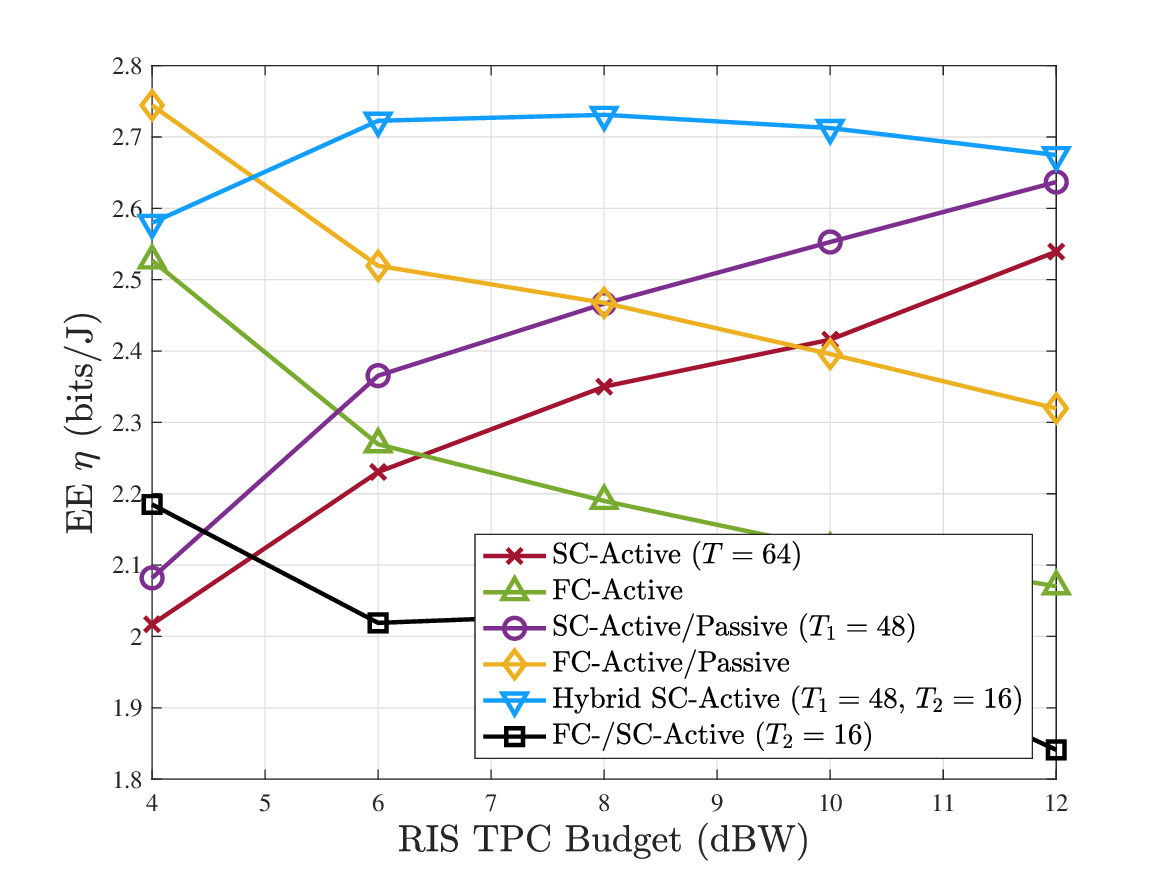} } 
	\subfloat[]{
	\label{subfig:6c}
	\includegraphics[scale = 0.3]{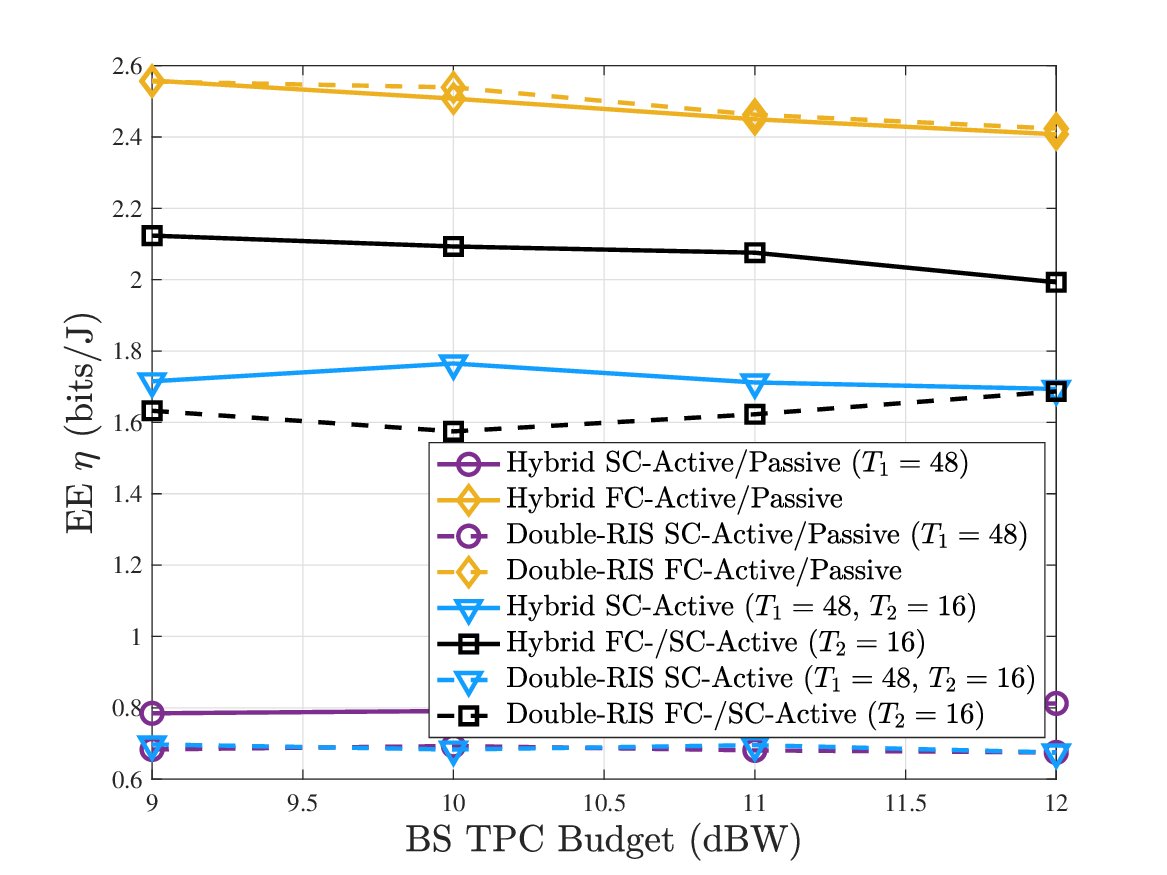} }

	\subfloat[]{
	\label{subfig:6d}
	\includegraphics[scale = 0.3]{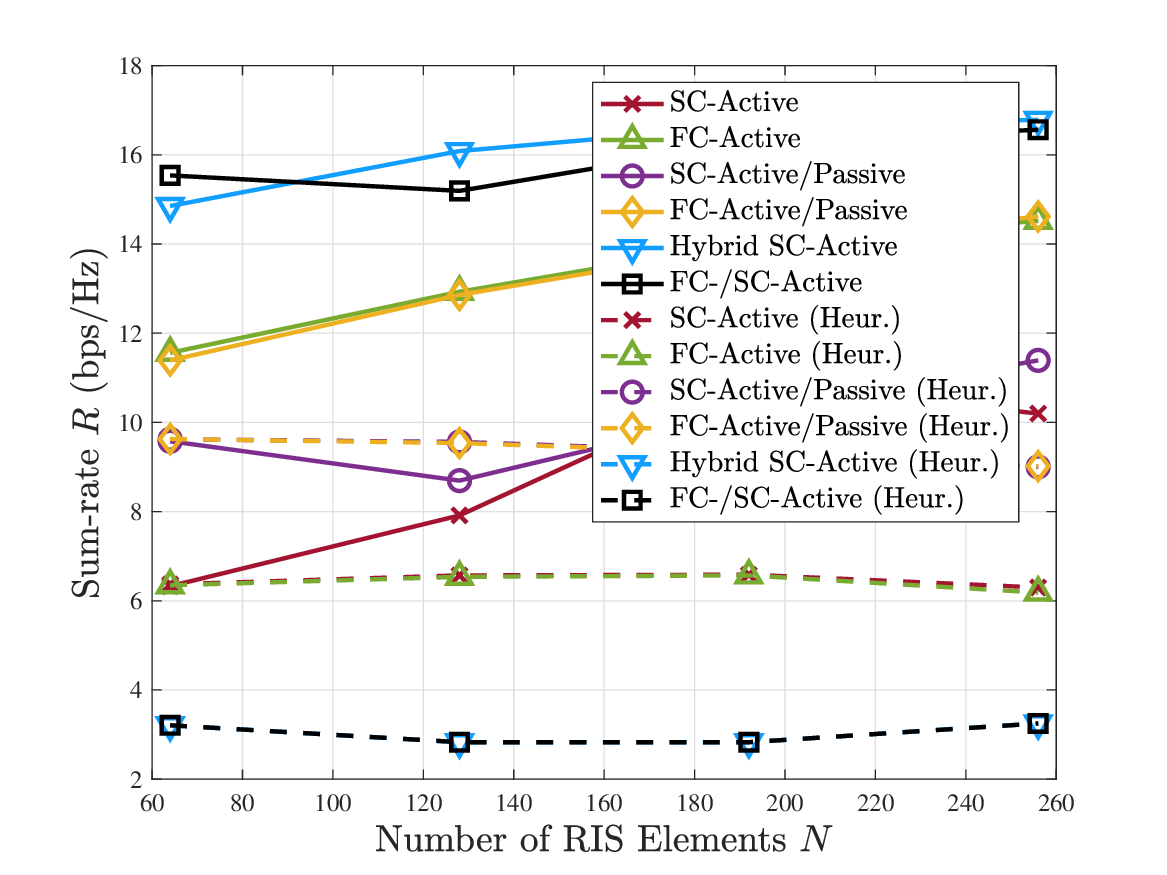} } 
	\subfloat[]{
	\label{subfig:6e}
	\includegraphics[scale = 0.3]{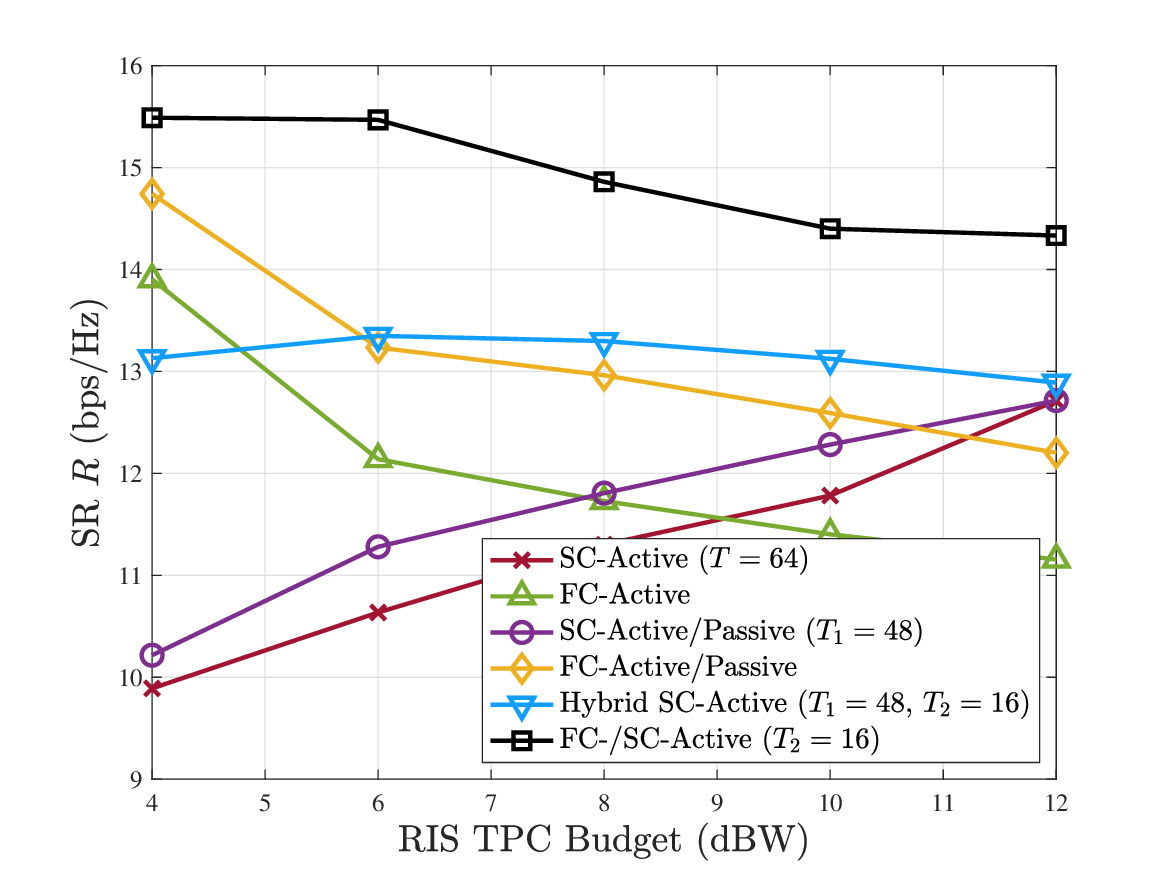} } 
	\subfloat[]{
	\label{subfig:6f}
	\includegraphics[scale = 0.3]{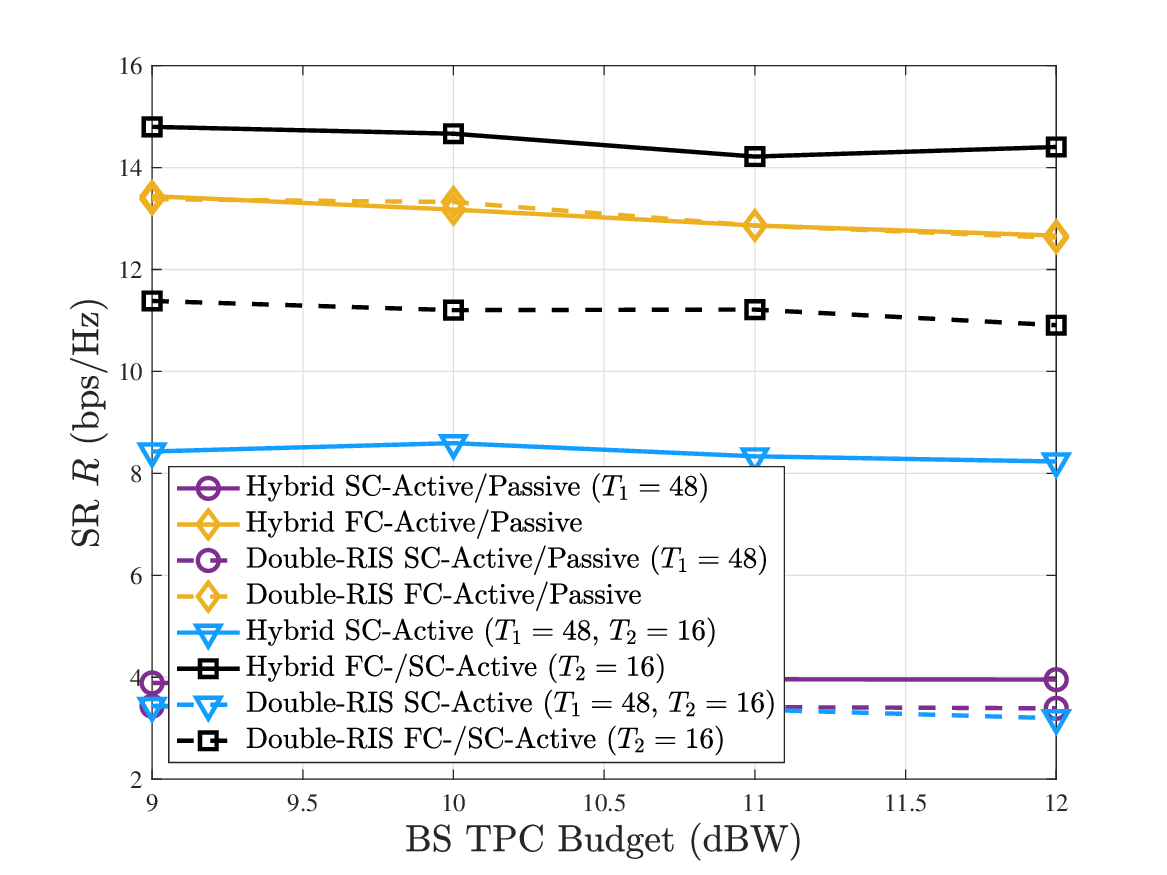} } 
\caption{EE $\eta$ (top) and SR $R$ (bottom) vs. $N=64:64:256$ for $a=0.5$, $L=4$, $D=300$ m (left), $P_{\text{RIS}}^{\max}=4:2:12$ dBW for $N=64$, $a=0.75$, $L=1$, $D = 300$ m (center), $P_{\text{BS}}^{\max}=9:12$ dBW for $N=64$, $a=0.75$, $L=1$, $D = 300$ m (right).}
\label{fig:6}
\end{figure*}

\textbf{Scaling with the Number of RIS Elements and Comparison with Heuristics:} The left column of Fig.~\ref{fig:6} shows that increasing the number of RIS elements leads to higher SR but may decrease the EE, except for designs with fixed, partition-based instead of elements-based number of amplifiers. We also compare the performance with a scenario where we utilize the zero-forcing (ZF) linear precoding scheme adapted to the overall direct channel matrix, $\mathbf{D}=\left[\mathbf{g}_1^{\dagger};\dots;\mathbf{g}_K^{\dagger}\right]\in\mathbb{C}^{K\times M}$, i.e., $\mathbf{W} = \alpha\mathbf{F}$, $\mathbf{F}=\mathbf{D}^{+}=\mathbf{D}^{\dagger}\left(\mathbf{D}\mathbf{D}^{\dagger}\right)^{-1}$, and $\alpha = \tilde{P}_{\text{BS}}^{\max}/\operatorname{Tr}\left(\mathbf{F}\mathbf{F}^{\dagger}\right)$, where $\mathbf{F}\in\mathbb{C}^{K\times M}$ is the non-normalized ZF precoder and $\alpha$ is a normalization parameter that ensures compliance with the TSP power constraint derived from the BS's TPC one. These heuristic designs perform worse than the optimal ones, with the SC-active/passive and FC-active/passive variants showing the best performance in terms of both EE and SR (outperforming the corresponding optimal SC-active/passive and SC-active designs when $N$ is small or moderate). On the other hand, they present lower complexity than joint transmit/RIS precoding optimization and do not face potential convergence issues due to alternating optimization. 

\textbf{Effect of RIS's Power Budget:} According to the central column of Fig.~\ref{fig:6}, hybrid SC-active RIS maintains the highest EE across power budgets, peaking at 8 dBW before declining. FC-active/passive RIS shows decreasing EE with budget increase, as the SR gains don't offset the higher power usage, while RIS variants with fixed count of shared amplifiers show rising EE due to the controlled power consumption.

\textbf{Effect of BS's Power Budget and Comparison with Double RIS Setups:} The right column of Fig.~\ref{fig:6} demonstrates that the hybrid RIS architectures outperform equivalent double RIS setups, with the active/active designs presenting larger relative gains. FC-active/passive RIS achieves the best EE, since with the considered settings its high SR offsets its large TPC. FC-active/SC-active RIS exhibits the best SR with increasing BS power budget, efficiently utilizing the additional transmit power via its enhanced signal control.
\section{Summary and Conclusions}\label{sec:6}
In this paper, we introduced novel hybrid RIS architectures with at least one SC-active RS and proposed a system model that captures the power combination and re-distribution attributed to power amplifier sharing, yet results in efficient optimization. Analytical studies have shown that these RIS structures achieve a significant fraction of the SNR of an equivalent active RIS, thus indicating the potential of significant EE and possibly SR gains for the proposed hybrid RIS designs over fundamental single-architecture RIS variants and FC-active RS-based hybrid RIS implementations, considering their TPC savings and the increase of the amplification noise with the number of power amplifiers. The analysis-derived projections have been confirmed via extensive studies based on computer simulations. According to the numerical results, the proposed hybrid SC-active and FC-active/SC-active RIS achieve the best EE and SR, respectively, in multi-user settings, while the proposed, power-efficient SC-active/passive RIS outperforms SC-active RIS, as expected, and even reaches the performance of hybrid SC-active RIS for large RIS TPC budget. In the future, we aim to optimize RIS elements' scheduling and model the TPC impact of the switches (e.g., diodes) used for flexible RIS configuration, as described in Sec.~\ref{subsec:2.2}.

\appendices
\renewcommand{\thesectiondis}[2]{\Alph{section}:}
\section{Proof of Lemma 3}\label{App:A}
The received signal is $y=\frac{\beta}{\sqrt{N_1}}\mathbf{f}_{1}^{\dagger}\bm{\Theta}_{1}^2\mathbf{g}_{1}ws+\mathbf{f}_{2}^{\dagger}\bm{\Theta}_{2}\mathbf{g}_{2}ws+\frac{\beta}{\sqrt{N_1}}\mathbf{f}_{1}^{\dagger}\bm{\Theta}_{1}^2\mathbf{z}_{1}+n$. Setting $\bm{\Psi}_1\triangleq\bm{\Theta}_1^2$, the SINR is given by
\begin{equation}\label{eq:App2}
\gamma_{a/p}=\underbrace{\frac{\left|\beta\mathbf{f}_{1}^{\dagger}\bm{\Psi}_{1}\mathbf{g}_{1}w\right|^{2}}{\beta^{2}\left\|\mathbf{f}_{1}^{\dagger}\bm{\Psi}_{1}\right\|^2 \delta_{1}^{2}+N_1\sigma^{2}}}_{\gamma_{a/p}^{(1)}}+\underbrace{\frac{N_1\left|\mathbf{f}_{2}^{\dagger}\bm{\Theta}_{2}\mathbf{g}_{2}w\right|^{2}}{\beta^{2}\left\|\mathbf{f}_{1}^{\dagger}\bm{\Psi}_{1}\right\|^2 \delta_{1}^{2}+N_1 \sigma^{2}}}_{\gamma_{a/p}^{(2)}}.
\end{equation}
To maximize $\gamma_{a/p}$, we initially seek to maximize $\gamma_{a/p}^{(1)}$: 
\begin{subequations}\label{eq:App3}
\begin{alignat}{2}
&&&\text{(SNR1-A): }\underset{w,\beta,\left\{\psi_{n,1}\right\}}{\max} \ \gamma_{a/p}^{(1)} \label{eq:App3a} \\
&&&\text{s.t.} \ \ \ \ \text{C7: }\left|w\right|^{2} \leq P_{t,a/p}^{\max}, \label{eq:App3b} \\
&&& \ \ \ \ \ \ \ \text{C8: }\beta^{2}\left\|\bm{\Psi}_{1}\mathbf{g}_{1}w\right\|^{2}+\beta^{2}N_1\delta_{1}^{2} \leq N_1 P_{r,1}^{\max}. \label{eq:App3c}
\end{alignat}
\end{subequations}
Using the Lagrange multipliers method, we obtain $w^{\star} = \sqrt{P_{t,a/p}^{\max}}$, $\psi_{n,1}^{\star} = \operatorname{arg}\left(f_{n,1}\right)-\operatorname{arg}\left(g_{n,1}\right)$, and 
\begin{equation}\label{eq:App4}
\beta^{\star} = \sqrt{\frac{N_1 P_{r,1}^{\max}}{P_{t,a/p}^{\max}\sum\limits_{n\in\mathcal{N}_{1}}\left|g_{n,1}\right|^{2}+N_{1}\delta_{1}^{2}}}.
\end{equation}

By substituting $w^{\star}$, $\beta^{\star}$, and $\theta_{n,1}^{\star}=\psi_{n,1}^{\star}/2$~\cite{SC3} into $\gamma_{a/p}^{(1)}$ and letting $N_{1}\rightarrow\infty$, such that, according to the law of large numbers, $\sum_{n\in\mathcal{N}_{1}}\left|f_{n,1}\right|\left|g_{n,1}\right| \rightarrow N_{1}\frac{\pi\varrho_{f_{1}}\varrho_{g_{1}}}{4}$, $\sum_{n\in\mathcal{N}_{1}}\left|f_{n,1}\right|^{2} \rightarrow N_{1}\varrho_{f_{1}}^{2}$, and $\sum_{n\in\mathcal{N}_{1}}\left|g_{n,1}\right|^{2} \rightarrow N_{1}\varrho_{g_{1}}^{2}$, we obtain
\begin{equation}\label{eq:App6}
\gamma_{a / p}^{(1)}\rightarrow N_1\frac{P_{t,a/p}^{\max} P_{r,1}^{\max}\pi^2\varrho_{f_{1}}^2\varrho_{g_{1}}^2}{16\left(P_{r,1}^{\max}\varrho_{f_{1}}^{2} \delta_{1}^{2}+ 4P_{t,a/p}^{\max}\varrho_{g_{1}}^{2}\sigma^2+4\delta_1^2\sigma^2\right)}.
\end{equation}
Next, given the optimal solution of (SNR1-A) and based on Eq.~(\ref{eq:App2}), we focus on maximizing the nominator of $\gamma_{a/p}^{(2)}$ subject to the UMC constraints of RS2: 
\begin{equation}\label{eq:App7}
    \text{(SNR1-B): }\underset{\left\{\theta_{n,2}\right\}}{\max} \ \left|\mathbf{f}_{2}^{\dagger}\bm{\Theta}_{2}\mathbf{g}_{2}w^{\star}\right|^{2} \ \text{s.t. C9: }0 \leq \theta_{n,2} \leq 2\pi.
\end{equation}
It is easy to verify that $\theta_{n,2}^{\star} = \operatorname{arg}\left(f_{n,2}\right)-\operatorname{arg}\left(g_{n,2}w^{\star}\right)$, $\forall n\in\mathcal{N}_{2}$~\cite{IRS1}. Therefore, by letting $N_{1}\rightarrow\infty$ and $N_{2}\rightarrow\infty$, such that $\sum_{n\in\mathcal{N}_{2}}\left|f_{n,2}\right|\left|g_{n,2}\right| \rightarrow N_{2}\frac{\pi\varrho_{f_{2}}\varrho_{g_{2}}}{4}$, we obtain
\begin{equation}\label{eq:App10}
    \gamma_{a/p}^{(2)} \rightarrow N_{2}^{2}\frac{P_{t,a/p}^{\max}\pi^{2}\varrho_{f_{2}}^{2}\varrho_{g_{2}}^{2}\left(P_{t,a/p}^{\max}\varrho_{g_{1}}^{2}+\delta_{1}^{2}\right)}{4\left(P_{r,1}^{\max}\delta_{1}^{2}\varrho_{f_{1}}^{2}+4P_{t,a/p}^{\max}\sigma^{2}\varrho_{g_{1}}^{2}+4\delta_{1}^{2}\sigma^{2}\right)}.
\end{equation}
Substituting Eqs.~(\ref{eq:App6}) and~(\ref{eq:App10}) into Eq.~(\ref{eq:App2}), we obtain Eq.~(\ref{eq:5}). This concludes the proof.  \qed

\section{Proof of Lemma 4}\label{App:B}
The received signal at the user is given by $y=h^{*}ws+\sum_{s\in\mathcal{S}}\mathbf{f}_{s}^{\dagger}\bm{\Phi}_{s}\mathbf{z}_{s}+n=\sum_{s\in\mathcal{S}}\frac{\beta_s}{\sqrt{N_s}}\mathbf{f}_{s}^{\dagger}\bm{\Theta}_{s}^2\mathbf{g}_{s}ws+\sum_{s\in\mathcal{S}}\frac{\beta_s}{\sqrt{N_s}}\mathbf{f}_{s}^{\dagger}\bm{\Theta}_{s}^2\mathbf{z}_{s}+n$. Thus, after manipulations,
\begin{equation}\label{eq:App12}
    \gamma_{a/a} = \sum_{s\in\mathcal{S}}\gamma_{a/a}^{(s)} = \sum_{s\in\mathcal{S}}\frac{\left|\beta_{s}\mathbf{f}_{s}^{\dagger}\bm{\Psi}_{s}\mathbf{g}_{s}w\right|^{2}}{\beta_{s}^{2}\left\|\mathbf{f}_{s}^{\dagger}\bm{\Psi}_{s}\right\|\delta_{s}^{2}+N_s\sigma^2}.
\end{equation}
The SNR maximization problem is formulated as follows: 
\begin{subequations}\label{eq:App13}
\begin{alignat}{2}
&&&\text{(SNR2): }\underset{w,\left\{\beta_{s}\right\},\left\{\psi_{n,s}\right\}}{\max} \ \gamma_{a/a}^{(s)} \label{eq:App13a} \\
&&&\text{s.t.} \ \ \ \ \text{C10: }\left|w\right|^{2} \leq P_{t,a/a}^{\max}, \label{eq:App13b} \\
&&& \ \ \ \ \ \ \ \text{C11: }\beta_{s}^{2}\left\|\bm{\Psi}_{s}\mathbf{g}_{s}w\right\|^{2}+\beta_{s}^{2}N_{s}\delta_{s}^{2} \leq N_s P_{r,s}^{\max}. \label{eq:App13c}
\end{alignat}
\end{subequations}
Using the Lagrange multipliers method to obtain the optimal solutions, substituting these expressions into $\gamma_{a/a}$, letting $N_{s}\rightarrow\infty$, and applying the law of large numbers, we get 
\begin{equation}\label{eq:App14}
\gamma_{a / a}^{(s)}\rightarrow N_s\frac{P_{t,a/a}^{\max} P_{r,s}^{\max}\pi^2\varrho_{f_{s}}^2\varrho_{g_{s}}^2}{16\left(P_{r,s}^{\max}\varrho_{f_{s}}^{2} \delta_{s}^{2}+ P_{t,a/a}^{\max}\varrho_{g_{s}}^{2}\sigma^2+\delta_s^2\sigma^2\right)}.
\end{equation}
Summing Eq.~(\ref{eq:App14}) over all $s\in\mathcal{S}$ and substituting $N_s=N/S$, we derive Eq.~(\ref{eq:18}). This concludes the proof. \qed


\bibliographystyle{IEEEtran}
\bibliography{refs2}

\end{document}